\documentclass[prl,preprint,epsfig,rotate,superscriptaddress,showpacs,amsmath,amsfonts,amssymb,floatfix]{revtex4-1}
\usepackage{graphicx}
\usepackage{dcolumn}
\usepackage{bm}
\usepackage{epstopdf}
\usepackage{dsfont}

\newcommand{\bra}[1]{\left\langle #1\right|}
\newcommand{\ket}[1]{\left| #1\right\rangle}
\newcommand{\ip}[2]{\left\langle #1 | #2\right\rangle}
\newcommand{\heff}{\hbar _{e}}
\newcommand{\eff}[1]{#1 _{e}}

\begin{document}

\title{Quantized
Adiabatic Transport in Momentum Space}

\author{Derek Y.H. Ho}
\affiliation{Department of Physics and Center for Computational
Science and Engineering, National University of Singapore, 117542,
Singapore}
\author{Jiangbin Gong}
\email{phygj@nus.edu.sg}
\affiliation{Department of Physics and Center for Computational
Science and Engineering, National University of Singapore, 117542,
Singapore}
\affiliation{NUS Graduate School for Integrative Sciences
and Engineering, Singapore 117597, Singapore}

\begin{abstract}

   Though topological aspects of energy bands are known to play a key role in quantum transport in solid-state systems, the implications
   of Floquet band topology for transport in momentum space (i.e., acceleration) are not explored so far.
   Using a ratchet accelerator model inspired by existing cold-atom experiments, here we characterize
   a class of extended Floquet bands of one-dimensional driven quantum systems
   by Chern numbers, reveal topological phase transitions therein,  and theoretically predict the quantization of
   adiabatic transport in momentum space.  Numerical results confirm
   our theory and indicate the feasibility of experimental studies.
   %evolution operators periodic in momentum can give rise to quantized transport in momentum space in exact analogy to QAPT. It is hoped that this work will %stimulate further discoveries of more intriguing momentum periodicity-related phenomena which parallel the well-known phenomena known due to spatial periodicity.
\end{abstract}

\date{\today}
\maketitle

%In both classical mechanics and quantum mechanics, position and momentum variables form a conjugate pair and can hence be
%treated on the same footing from a phase space perspective. The real physical world, however, does not have position-momentum
%symmetry. For example, energy bands of a solid
%are formed because electrons move in a potential periodic in position but not in momentum. Therefore,
%mapping quantum transport phenomena from position space to momentum space is nontrivial if possible:
%new physical insights may emerge and unforeseen methodologies or quantum simulation strategies may be developed.  Consider Anderson localization as
%a seminal result of quantum transport in position space. Its analog in momentum space is the so-called ``dynamical localization" \cite{Fishman1984DynLocaliztnPRA}.
%This mapping has led to fruitful studies of the universality of
%Anderson transition in driven cold-atom systems \cite{Chabe2008PRL,*Garreau2012PRL}.  As a second example, ratchet transport \cite{Dana2010Ratchets_IJBC,*Dana2008experimentalratchetPRL}, namely, directed transport
%under a zero mean force in position space, can be mapped to momentum space as well, leading to the finding of ratchet accelerator (RA)
%albeit time-reversal symmetry \cite{Gong2004ratchetPRE,*Gong2006ratchetPRL}.

%Add references on Floquet topological insulators, PRL on Floquet state anholonomy,

In both classical mechanics and quantum mechanics, position and momentum variables form a conjugate pair and can hence be
treated on the same footing from a phase space perspective. The real physical world, however, does not have position-momentum
symmetry. For example, energy bands of a solid
are formed because electronic Hamiltonians are periodic in position but not in momentum. Due to such unequal roles of position and momentum,
the mapping of quantum transport phenomena from position space to momentum space is typically nontrivial but, where possible, may lead to
important insights and unforeseen opportunities. Anderson localization, for instance, was first discovered as
a seminal result of quantum transport in position space. Its analog in momentum space was later found to be behind the intriguing phenomenon of  ``dynamical localization" \cite{Fishman1984,Casatibook1995}.
This mapping has stimulated fruitful studies of
Anderson transition in driven cold-atom systems \cite{Chabe2008PRL}.  As a second example, ratchet transport, namely, directed transport
under a zero mean force in position space, has been mapped to momentum space as well, leading to the finding of ratchet accelerators (RA)
%albeit time-reversal symmetry
\cite{Gong2004ratchetPRE, Lundh2005PRL, Dana2008}.

Energy-band topology is of fundamental interest to studies of quantum transport in condensed-matter physics. %\cite{Raizen1995KRrealizationPRL,Kanem2007observationPRL,Sadgrove2007rectifiedPRL,*Dana2008experimentalratchetPRL,*Ryu2006highorderPRL,Talukdar2010subfourierPRL,Hoogerland2011LoschmidtPRE}, as manifested in the literature of
%quantum Hall effect \cite{TKNN1984QHEPRL}, quantized adiabatic pumping \cite{Thouless1983QAPTPRB,*NiuThouless1984QAPTJPhysA},
%topological insulators \cite{Kane2011TopoInsulatorPhysWorld,*Kane2010TopoInsulatorsRMP} etc.
The issue to be addressed here is whether transport in momentum space (i.e., acceleration) can be connected with band structures due to
momentum space periodicity.
At first glance this sounds impossible because, with the (non-relativistic) kinetic energy being a
quadratic function of momentum, a realistic Hamiltonian is never a periodic function of momentum.
However, for systems periodically driven by impulsive fields, the Floquet operator can still be periodic in momentum, thus forming Floquet (quasi-energy) bands.
The topology of such Floquet bands then provides a useful tool in characterizing topological phase transitions in driven quantum systems \cite{Kitagawa2010TopoCharacterizationPRB}.
How Floquet-band topology is manifested in acceleration then becomes an intriguing question. %\cite{Galitski2011FloquetTopoInsulatorNaturePhys}

Using a one-dimensional RA model inspired by existing cold-atom experiments, we show in this work that the Floquet bands,
defined on a 2-torus (formed by one Bloch phase and one experimentally tunable parameter) may be characterized by Chern numbers.
We then theoretically show that adiabatic transport in \emph{momentum space} can be
quantized according to these topological numbers. Unlike quantized adiabatic pumping in position space \cite{Thouless1983QAPTPRB},
there does not exist a general (system-independent) flux operator in momentum space and hence
the found quantization is about a quantized net change of momentum expectation value, rather than a pumping of particles through a cross section.
Finally, though adiabatic manipulation of Floquet states is often
considered to be subtle \cite{Kohn1997AdiabaticLimitPRA,Dranov1998DiscreteAdiabaticJMathPhys,Tanaka2011DiscreteAdiabaticPhySocJpn},
our numerical results confirm our theory and indicate that quantized
momentum space transport can be observed in a wide regime,  by simply scanning one system parameter in a relatively
small number of discretized steps.  Breakdown of quantization in momentum space transport may then be considered
as a diagnostic tool for detecting non-adiabatic effects in Floquet state manipulation.

To physically realize the main physics we need three ingredients: momentum space periodicity, an experimentally tunable periodic parameter and well-gapped Floquet bands.
To be specific we
consider a RA model \cite{WangPRE2008}, which is based on an atom-optics realization of a double-kicked rotor system \cite{MonteiroDoubleKick}. The RA Hamiltonian is given by
$H = \frac{{p}^{2}}{2} + K \cos({q}+\alpha) \sum_{n} \delta (t-nT) + \: K\cos({q})\sum_{n} \delta(t-nT-T_{0})$,
with a corresponding Floquet propagator \cite{WangPRE2008} $\hat{U} (\alpha) = e ^{-i(T-T_{0})({p} ^{2} / 2 \hbar)}e^{-i(K / \hbar)\mathrm{cos}({q})}e^{-iT_{0}({p} ^{2} / 2 \hbar)} e^{-i (K / \hbar)\mathrm{cos}({q}+\alpha)} $,
where all quantities are properly scaled and hence in dimensionless units, ${q}$ and ${p}$ are canonical coordinate and momentum operators for cold atoms,
 and the $\delta$ kicks are of period $T$, experimentally
 implemented by two optical-lattice potentials mutually phase-shifted by $\alpha$, with equal strength $K$, and a time lag
 $T_{0} < T $ \cite{MonteiroDoubleKick}. Note that later $\alpha$ will be adiabatically tuned.
 Because the potential function is periodic in $q$, momentum eigenstates take eigenvalues  $m \hbar + \beta \hbar $,  where $\beta\in [0,1] $ is a conserved quasi-momentum variable and $m$ is an integer. To yield a Floquet operator periodic in momentum despite the $p^2/2$ term in the Hamiltonian,
 we set $\beta=0$, which may be approximately implemented by considering a
 Bose-Einstein condensate whose coherence width spans across many optical lattice constants \cite{Dana2008,Ryu2006highorderPRL,Talukdar2010subfourierPRL}.
 Effects of nonzero $\beta$ values will be discussed in Appendix \cite{supple}.
 If we now impose the quantum resonance condition $ T \hbar = 4\pi $ that has been one experimental subject \cite{Raizen1995KRrealizationPRL,Kanem2007observationPRL,Dana2008,Ryu2006highorderPRL,Talukdar2010subfourierPRL,Hoogerland2011LoschmidtPRE}, we obtain an ``on-resonance double-kicked-rotor model'' \cite{Gong2008proposalPRA}, with the Floquet propagator reducing to
\begin{eqnarray} \label{beta_rotor}
\hat{U} ^{}_{R} (\alpha) &=& e^{i\frac{{\eff{p}}^{2}}{2\heff}}e^{-i\frac{\eff{K}}{\heff} \mathrm{cos}({q})}e^{-i\frac{{\eff{p}}^{2}}{2\heff}} e^{-i\frac{\eff{K}}{\heff} \mathrm{cos}({q}+\alpha)},
\end{eqnarray}
where $\heff \equiv \hbar T_{0}$ is an effective Planck constant, $\eff{K} \equiv K T_{0} $, and a rescaled momentum operator
${\eff{p}} \equiv T_0 p$. From here on, momentum exclusively refers to $\eff{p}$ and
we denote momentum eigenstates by $|m\rangle$, which has eigenvalue $m\heff$ and is periodic in $q$ with period $2\pi$.
%In actual experiments, a continuous spread of $\beta$ values over a finite range is always present. The practical effects of this will be considered later. For now, %however, we restrict ourselves to the idealization $\beta = 0$ corresponding to a genuine rotor. In this scenario, the Floquet operator simplifies to become
%\begin{eqnarray} \label{beta_zero_rotor}
%\hat{U} ^{(\beta=0)}_{R} (\alpha) &=& e^{+i((\hat{m}\heff)^{2} / 2\heff)} e^{-i(\eff{K}/\heff)\mathrm{cos}(\hat{q})}e^{-i((\hat{n}\heff)^{2}/2\heff)}\nonumber \\
%&& \times \: e^{-i(\eff{K}/ \heff)\mathrm{cos}(\hat{q}+\alpha)}.
%\end{eqnarray}
 Equation (\ref{beta_rotor}) indicates that if $\heff = 2\pi M/N$, with $M$ and $N$ being integers, then
the Floquet operator $\hat{U}^{}_{R}$ is perfectly periodic in momentum space with a period of $N \heff$. According to Bloch's theorem, this
 momentum space periodicity leads to Floquet bands. Indeed, in the case of $\alpha=0$, the corresponding Floquet band structure \cite{Gong2008proposalPRA}
resembles Hofstadter's butterfly \cite{Hofstadter1976ButterflyPRB}, with $\heff$ here identified as an analog of the magnetic flux.
It is such a remarkable resemblance (which also hints desirable band-gap features)
% studies of quantum hall effects. It is due to this similarity of band structures that we might expect the system to display phenomena related to those observed in %quantum hall effect studies, such as the QAPT \cite{Thouless1983QAPTPRB}.
that motivated us to connect the
topological aspects of the Floquet bands with momentum-space transport.

\begin{figure}[h!]
\begin{center}$
\begin{array}{c}
%\centering
\includegraphics[width=60mm, height=52mm] {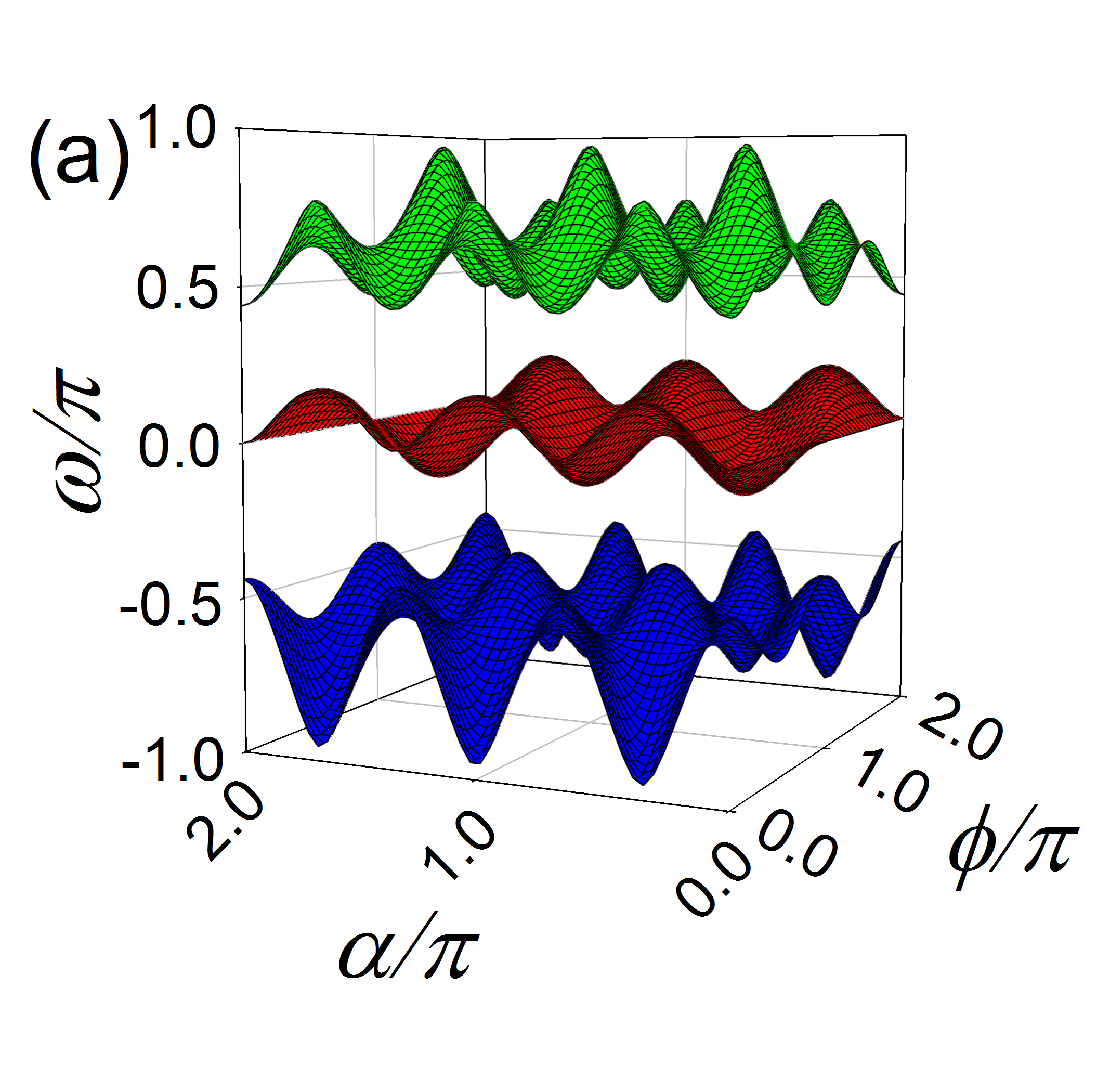}  \hspace{-0.cm}
\includegraphics[width=60mm, height=52mm] {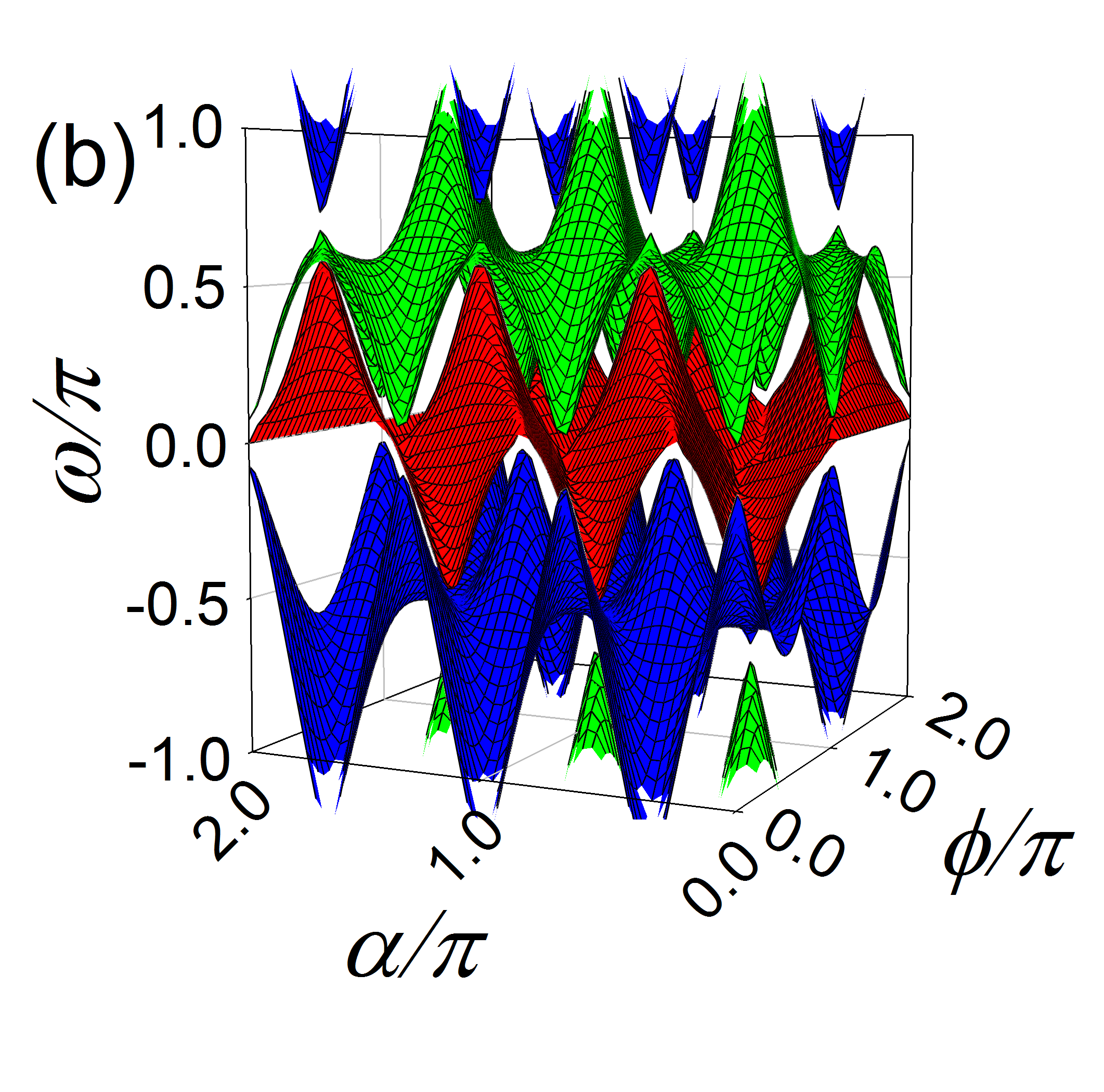}
\end{array}$
\end{center}
\caption{(color online) Floquet eigenphase $\omega_{n}(\phi,\alpha)$ vs $\phi$ and $\alpha$ for (a) $ \eff{K} = 3\heff $ and (b) $ \eff{K} = 4 \heff $.}
 %(c) $ \eff{K} = 5\heff$. It can be seen that the bands deform as $\eff{K}$ increases, developing cone-like structures of increasing height until the cones collide %at the tips at a critical $ \eff{K} $ and then the Chern numbers change. After the transition, the bands further deform with $K_e$.}
 \label{bands}
\end{figure}

The Floquet band structure may be characterized by topological Chern numbers, provided that
the bands are defined on a 2-torus.  To that end, we supplement the Bloch phase parameter
in momentum space with the periodic parameter $\alpha$. This procedure somewhat resonates with recent efforts in identifying
analogs of quantum Hall effect in one-dimensional systems \cite{Beenakker1DQHE2011PRB,*ChenShuEdgestates2011arXiv,*FengZ2insulators2012PRA,*KrausTopologicalQuasicrystals2011arXiv}.
The eigenstate-eigenvalue problem for $\hat{U} ^{}_{R}$ now becomes $
\hat{U}_{R} (\alpha) \ket{\psi _{n} (\phi,\alpha)} = e^{i\omega_{n}(\phi,\alpha)}\ket{\psi _{n} (\phi,\alpha)}$,
where $\phi \in [0,2\pi)$ is the Bloch phase in momentum space
and $\omega_{n}(\phi,\alpha)$ is the eigenphase of $\hat{U}_{R} (\alpha)$. For a fixed pair of $\alpha$ and $\phi$, $N$ eigenphases can be obtained, and scanning $(\phi,\alpha)$ over $[0,2\pi)\times [0,2\pi)$
forms {\it extended} Floquet bands on 2-torus, so-named to distinguish them from the common bands involving only the Bloch phase parameter
$\phi$. As a result $n$ ($1\le n \le N$) becomes an index of such bands (with $\heff=2\pi M/N$).
The eigenstates $\ket{\psi _{n} (\phi,\alpha)}$ are chosen such that they are locally single-valued functions of $\alpha$ and $\phi$,
and periodic functions of $\phi$.  Two computational examples of the Floquet bands $\omega_{n}(\phi,\alpha)$
are depicted in Fig.~\ref{bands} for $N=3$.  It is seen that as the driving strength $K$ varies,
the landscape of the bands changes.  Other calculations show that if $\heff = 2\pi M/N$ with even $N$, then
there will be two (extended) bands touching each other. For simplicity, here we consider only odd $N$ so that only accidental band collisions occur.
%It is also found that when $\heff = 2\pi M/N$ with even $N$, then two (extended) bands
%are always touching. To simplify the matter we choose odd $N$ so that only accidental band collision occurs.

%\begin{figure}
%\includegraphics[width=8.4 cm]{Fig1.PNG}
%\caption{The 3 extended bands obtained for the case of $\heff=2\pi /3$, with %$\eff{K}=3\heff$. }
%\label{bands}
%\end{figure}

%Quantitatively, if the bands are not touching each other,
%then each band can be associated with a Chern number, $ C_{n}$, namely,
%$C_{n} = \frac{i}{2\pi} \oint  \bra{\bar{\psi} _{n} ( \Gamma )} \frac{\partial}{\partial \Gamma} \ket{\bar{\psi}_{n} (\Gamma) } d \Gamma$,
%where $ \Gamma \equiv (\phi, \alpha) $, $\ket{\bar{\psi}_{n} (\Gamma) } \equiv \sum _{m=1} ^{N} \ket{m}\ip{m}{\psi _{n} (\Gamma)}$ is normalized
%on $N$ momentum states, and the circulation integral covers the entire 2-torus.
%

In Fig. \ref{cherntable}  we present the Chern numbers (see its surface integral expression below and see \cite{supple} for computational details)
for a 3-band case. % as a function of $K_e$.
It is seen that at some isolated critical values of $K_e$, the Chern numbers jump, signaling the presence of
toplogical phase transitions in our RA model. Indeed, the Chern numbers are invariant integers with respect to smooth
deformation of the bands and only change discontinuously due to band collisions.
Note that mainly
 in the context of quantum-classical correspondence in classically chaotic systems,
 Ref.~\cite{Leboeuf1992TopologicalAspectsChaos} (see also Ref.~\cite{DanaExtendednLocalized1995PRE,*DanaBandDistributions1998PRL})
 formally studied the topological aspects of
Floquet bands defined on a 2-torus formed by Bloch phases in both position space and momentum space.
Our study is much different because (i) this work is based on
an explicit physical implementation of momentum-space periodicity, (ii) here
the Bloch phase in position space is fixed at $\beta=0$ in theory (so as to obtain Hofstadter's butterfly Floquet spectrum),
and (iii)
 %In these early studies %\cite{Leboeuf1992TopologicalAspectsChaos,DanaExtendednLocalized1995PRE,*DanaBandDistributions1998PRL,*DanaBandHusimi1998PRE,DanaVortex2002JPhysA}
  our Floquet bands are defined on a 2-torus that involves one experimental parameter, a key starting point for our theory below.
\begin{figure}
\includegraphics[trim=1cm 7cm 1cm 5cm,clip=true,height=5.0cm ,width=!]{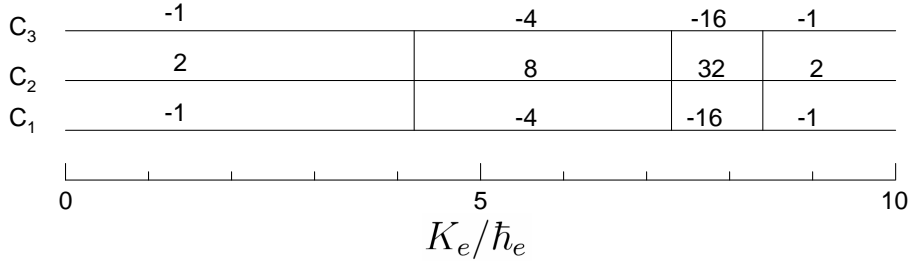}
\caption{The Chern numbers of the 3 bands vs $\eff{K} / \heff$, for $\heff=2\pi /3$.
%At critical values of $\eff{K}$, the extended bands collide and Chern numbers change. Note also that the Chern numbers can undergo big changes across a critical %point.
For results of a 7-band case, see Appendix \cite{supple}.}
\label{cherntable}
\end{figure}

We now seek the implications of the Floquet-band Chern numbers by considering an adiabatic cycle during which
$\alpha$ increases from $0$ to $2\pi$.
We first construct an initial state of the following form,
\begin{equation} \label{initialstate}
\ket{\Psi_{n} (\alpha=0) } = \frac{1}{2\pi} \int ^{2\pi} _{0} d\phi \ket{\psi _{n} (\phi,\alpha=0)},
\end{equation}
which represents an equal-weight superposition of all the Floquet eigenstates of band $n$ with $\alpha=0$. This coherent superposition state, which can be
interpreted as a Wannier function in momentum space,
uniformly samples all the Bloch eigenstates with different values of $\phi$ (but all with  $\alpha=0$),
with a profile localized in the momentum space.
Indeed,  each
eigenstate $\ket{\psi _{n} (\phi,\alpha=0)}$ is infinitely extended in momentum space, but $\ket{\Psi_{n} (\alpha=0)}$ is normalized to unity (localized in momentum space with Gaussian-like tails in all the cases we studied). It is worth noting that because each eigenstate $\ket{\psi _{n} (\phi,\alpha=0)} $ is defined only up to a global phase, one is free to choose an overall phase convention of $ \ket{\psi _{n} (\phi,\alpha)} $ such that the superposition state in momentum space tends to be well-localized, thus making experimental preparation of the initial state easier. Such states can be highly localized so long as $K$ is not too large. For example, the shown superposition state in Fig. 3(a) mainly occupies one momentum eigenstate, with small weights distributed over only a few nearby components.
Given previous experiments
where momentum superposition states in the same context were prepared \cite{Dana2008},
states as shown in Fig. 3(a) should be reachable in experiments.

%Indeed,  each
%eigenstate $\ket{\psi _{n} (\phi,\alpha=0)}$ is infinitely extended in momentum space, but $\ket{\Psi_{n} (\alpha=0)}$ is
%normalized to unity (localized in momentum space with Gaussian-like tails in all the cases we studied).
%Interestingly, states constructed in this manner can be highly localized so long as $K$ is not too large, with one such example displayed in Fig. 3(a).
\begin{figure}[h!]
\resizebox{9.4cm}{2.5cm}{
\begin{tabular}{c}
\includegraphics{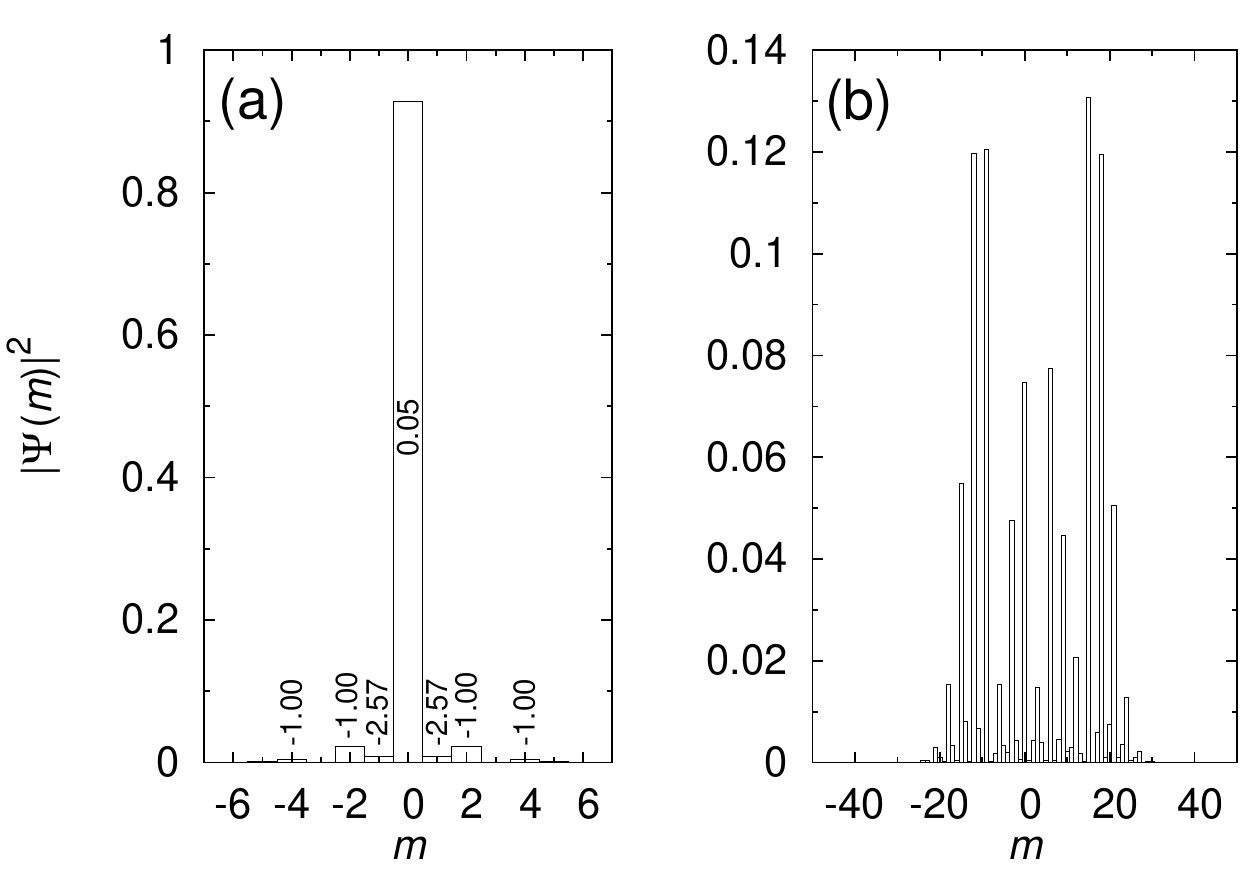}
\end{tabular}
}
\caption{\small{Momentum distribution $\left| \ip{m}{\Psi_{3} (\alpha = 0)} \right|^{2} \equiv \left| \Psi (m) \right| ^{2} $ [see Eq.~(2)] with $ \eff{K} = 2 \heff $ for $t=0$ in (a) and after a 100-period adiabatic cycle in (b).  Numbers shown in (a) are the phases of each
momentum component.}}
\label{p_profiles}
\end{figure}
%It is worth noting that because each eigenstate $\ket{\psi _{n} (\phi,\alpha=0)} $ is defined only up to a global phase,
%one is free to choose an overall phase convention of $ \ket{\psi _{n} (\phi,\alpha)} $ such that the superposition state
%in momentum space tends to be well-localized, thus making experimental preparation of the initial state easier. For instance, the shown superposition state in Fig. %3(a) mainly occupies one momentum eigenstate, with small
%weights distributed over few other nearby components.

Consider then an adiabatic change in $\alpha$, through a discretized protocol $\alpha_s =  2\pi s/s_f $ for $(s-1)T \leq t < sT$, so that
$\alpha$ completes one adiabatic cycle at $t=s_f T$ \cite{sadgrove}.   Assuming adiabatic following of the Floquet states, the state evolved from
$\ket{\Psi_{n} (\alpha=0)}$ [see Eq.~(\ref{initialstate})] should stay as a superposition state at $t=(sT)^{-}$, with each component
still being the eigenstate of $\hat{U}_{R} (\alpha)$, with $\alpha=\alpha_s$.  That is, at $ t=(sT)^{-}$, under adiabatic approximation the associated
time-evolving state of the system becomes
\begin{equation}
\ket{\Psi_{n} (\alpha_ {s}) } = \frac{1}{2\pi} \int ^{2\pi} _{0} d\phi \ket{\psi _{n} (\phi,\alpha _{s})} e^{i\theta (\phi,\alpha_{s})},
\end{equation}
where $\theta(\phi,\alpha_s)$ is the sum of a dynamical phase and a geometrical phase accumulated
by the component starting from $\ket{\psi _{n} (\phi,\alpha=0)}$. As a consequence of choosing $\ket{\psi _{n} (\phi,\alpha=0)}$ to be a periodic function
of $\phi$, $\theta(\phi,\alpha_s)$ is also necessarily periodic in $\phi$.

Next we evaluate  $\langle  \eff{p}(s)\rangle\equiv \langle \Psi_{n}(\alpha_s)| \eff{p} | \Psi_{n}(\alpha_s)\rangle$, namely,
the momentum expectation value on the state $ \ket{\Psi_{n} (\alpha _{s})} $.    To proceed
we first write Floquet eigenstates $\ket{\psi_{n} (\phi,\alpha)}$
as the product of two parts using the Bloch theorem, namely,
$\ket{\psi_{n} (\phi, \alpha)} = \hat{X}(\phi) \ket{u_{n} (\phi ,\alpha)}$,
with $ \hat{X} (\phi) \equiv e^{i{\eff{p}}\phi / N\heff }$.  We then define $|\bar{u}_n(\phi,\alpha)\rangle\equiv
\sum_{m=1}^{N} |m\rangle\langle m \ket{u_{n} (\phi ,\alpha)}$.
It can be shown that
the $N$-element state $|\bar{u}_n(\phi,\alpha)\rangle $ is an eigenstate of the following $N \times N$ reduced Floquet matrix
\begin{equation}
\langle m|\bar{U}(\phi,\alpha)|m'\rangle \equiv \sum^{\infty} _{l = - \infty} \langle m| \hat{X}^{\dagger} (\phi) \hat{U}_{R} (\alpha) \hat{X}(\phi)| m'+lN\rangle,
\end{equation}
with $ m, m' = 1, \cdots , N $.
 %$ [u_{n} (\phi ,\alpha)]^{\dagger} $ refers to the row matrix which is the adjoint of  column matrix $[u _{n} (\phi ,\alpha)]$. In what follows, the appearance of %such square-bracketed symbols adjacent to one another is to be understood to mean matrix multiplication.
Lengthy but straightforward calculations \cite{supple} then yield a compact expression for $\langle  \eff{p}\rangle$, i.e.,
$\langle  \eff{p}\rangle  = \frac{N}{2\pi}\int ^{2\pi} _{0} d \phi  \langle \bar{u}_{n} (\phi ,\alpha _{s})| i \heff \frac{\partial}{\partial \phi} |\bar{u} _{n} (\phi ,\alpha _{s})\rangle$.
To compare this expectation value with its preceding value for $\alpha=\alpha_{s-1}$,  we consider a first-order perturbation theory so as to
rewrite $|\bar{u} _{n} (\phi ,\alpha _{s})\rangle$ in terms of $|\bar{u} _{n} (\phi ,\alpha _{s-1})\rangle$.  Specifically, to the first-order of
$\delta\alpha\equiv \alpha_s-\alpha_{s-1}= 2\pi/s_f$, we have \cite{supple}
\begin{eqnarray} \label{1st_order_state}
&& \ket{\bar{u} _{n} (\phi ,\alpha _{s})}
= \ket{\bar{u} _{n} (\phi ,\alpha _{s-1})} \nonumber \\
&& + \: \delta\alpha \sum_{n'=1,\neq n} ^{N} \frac{\langle \bar{u} _{n'} (\phi ,\alpha _{s-1})| \frac{\partial \bar{U}(\phi,\alpha _{s-1})}{\partial \alpha_{s-1}}
 | \bar{u} _{n} (\phi ,\alpha _{s-1})\rangle }{e^{i \omega _{n} (\phi,\alpha_{s-1})}-e^{i \omega _{n'} (\phi,\alpha_{s-1})}}  \nonumber \\
&& \times \: |\bar{u} _{n'} (\phi ,\alpha _{s-1})\rangle.
\end{eqnarray}
The change in momentum expectation value (denoted by $\delta  \langle \eff{p} \rangle_s$) over the period from $t=(s-1)T$ to $(sT)^{-}$ can now be calculated
to the first order of $\delta\alpha$,
\begin{eqnarray}
\delta  \langle \eff{p} \rangle_s &=& \frac{1}{2\pi}N\int ^{2\pi} _{0} d \phi \Big\{ \langle \bar{u}_{n} (\phi ,\alpha _{s})| i \heff \frac{\partial}{\partial \phi} |\bar{u} _{n} (\phi ,\alpha _{s})\rangle  \nonumber  \\
&& -\: \langle \bar{u}_{n} (\phi ,\alpha _{s-1})| i \heff \frac{\partial}{\partial \phi} |\bar{u} _{n} (\phi ,\alpha _{s-1})\rangle \Big\}.
\label{small_p_change}
\end{eqnarray}
Further substituting Eq.~\eqref{1st_order_state} into Eq.~\eqref{small_p_change} yields
\begin{equation}
\delta \langle {\eff{p}} \rangle _{s} = -\frac{N}{2\pi} \int _{0} ^{2\pi} d\phi \: B _{n} (\phi,\alpha_s) \: \delta \alpha ,
\end{equation}
where $ B _{n} (\phi,\alpha)$ is identified as the Berry curvature
\begin{eqnarray}
 B _{n} (\phi,\alpha) =  i \sum _{n'=1,\neq n} ^{N} \Bigg\{ \frac{\langle \bar{u}_{n}|\frac{\partial \bar{U}^{\dagger}}{\partial  \phi}|\bar{u}_{n'}\rangle\langle \bar{u}_{n'}
|\frac{\partial \bar{U}}{\partial \alpha}|\bar{u}_{n}\rangle}{|e^{i \omega _{n}}-e^{i \omega _{n'}}|^{2}} - \mathrm{c.c} \Bigg\}, \nonumber \\
\end{eqnarray}
with the explicit dependences of $|\bar{u}_{n}(\phi ,\alpha)\rangle$, $\omega _{n} (\phi,\alpha)$ and $\bar{U}(\phi,\alpha)$ on $\alpha$ and $\phi$ all
suppressed for brevity.

The total change in momentum expectation value from $t=0$ to $(sT)^{-}$ is denoted by $ \Delta_{\eff{p}}(s)$.
Because $ \Delta_{\eff{p}}(s) = \sum _{s'=1} ^{s} \delta \langle {\eff{p}} \rangle _{s'} $,
we have
\begin{eqnarray}
\Delta_{\eff{p}}(s_{f})   &=& -  M \int _{0}^{2\pi} d\phi \int _{0}^{2\pi} d\alpha \: B _{n} (\phi,\alpha) \nonumber \\
&= &- 2 \pi M C_{n},
\label{finaleq}
\end{eqnarray}
where we have used $\heff = 2\pi M/N$ and $ C _{n} $ is exactly the Chern number of the $n$th Floquet band, whose surface integral
expression is $C_{n} = \frac{1}{2 \pi} \int _{0} ^{2\pi} d\phi \int _{0} ^{2\pi} d\alpha \: B _{n} (\phi,\alpha)$ \cite{supple}.
Thus, Eq.~(\ref{finaleq}) reveals that the net change in the momentum expectation value
over one adiabatic cycle of $\alpha$ (starting from state $\ket{\Psi_{n} (\alpha=0)}$) is quantized: it should be
proportional to the Chern number of the $n$th extended Floquet band.  This is our central theoretical result.

\begin{figure}[h!]
\begin{tabular}{c}
\rotatebox{-90}{
\resizebox{6.6cm}{!}{
\includegraphics{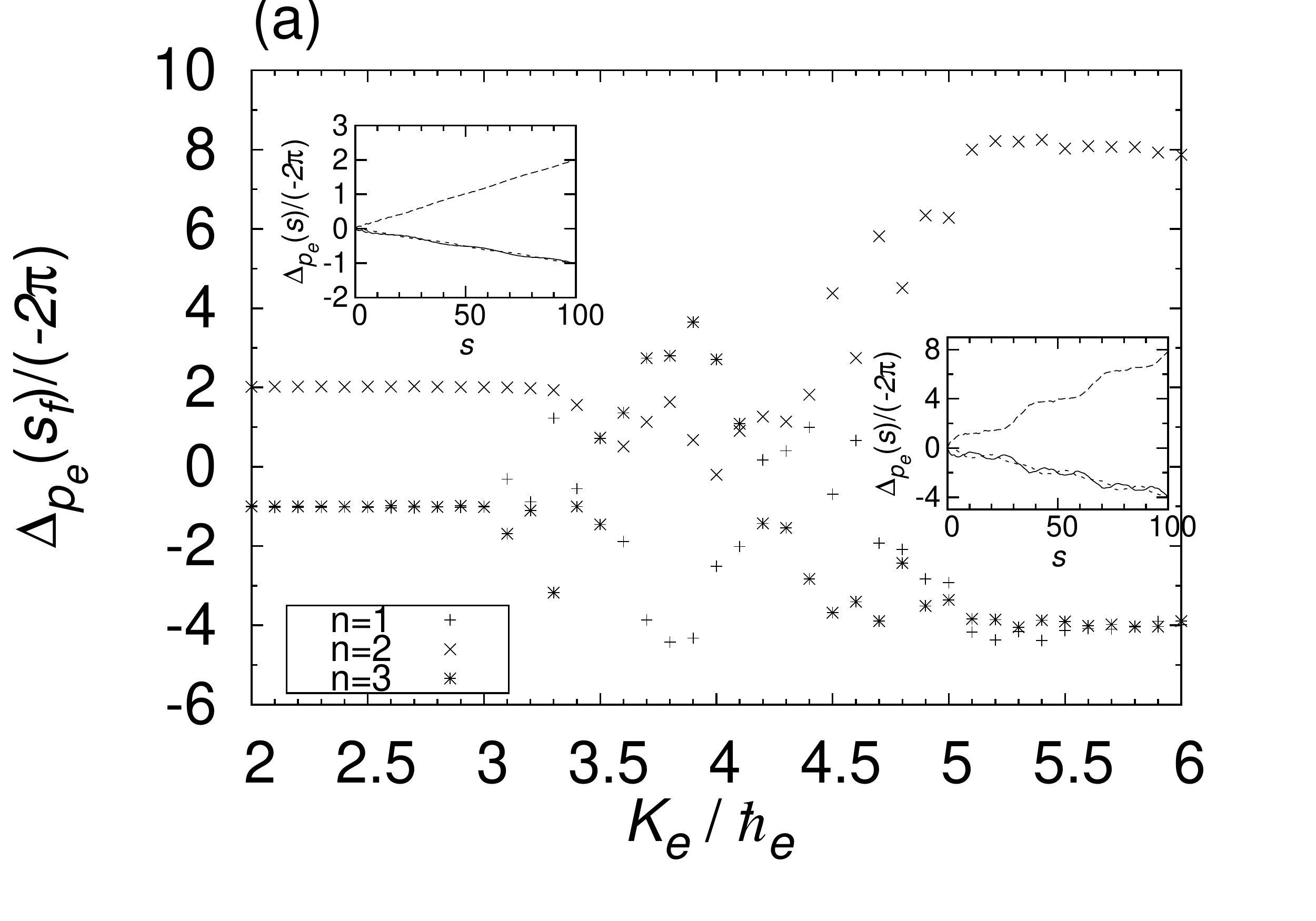} }
} \vspace{-1cm} \\
\rotatebox{-90}{
\resizebox{6.6cm}{!}{
\includegraphics{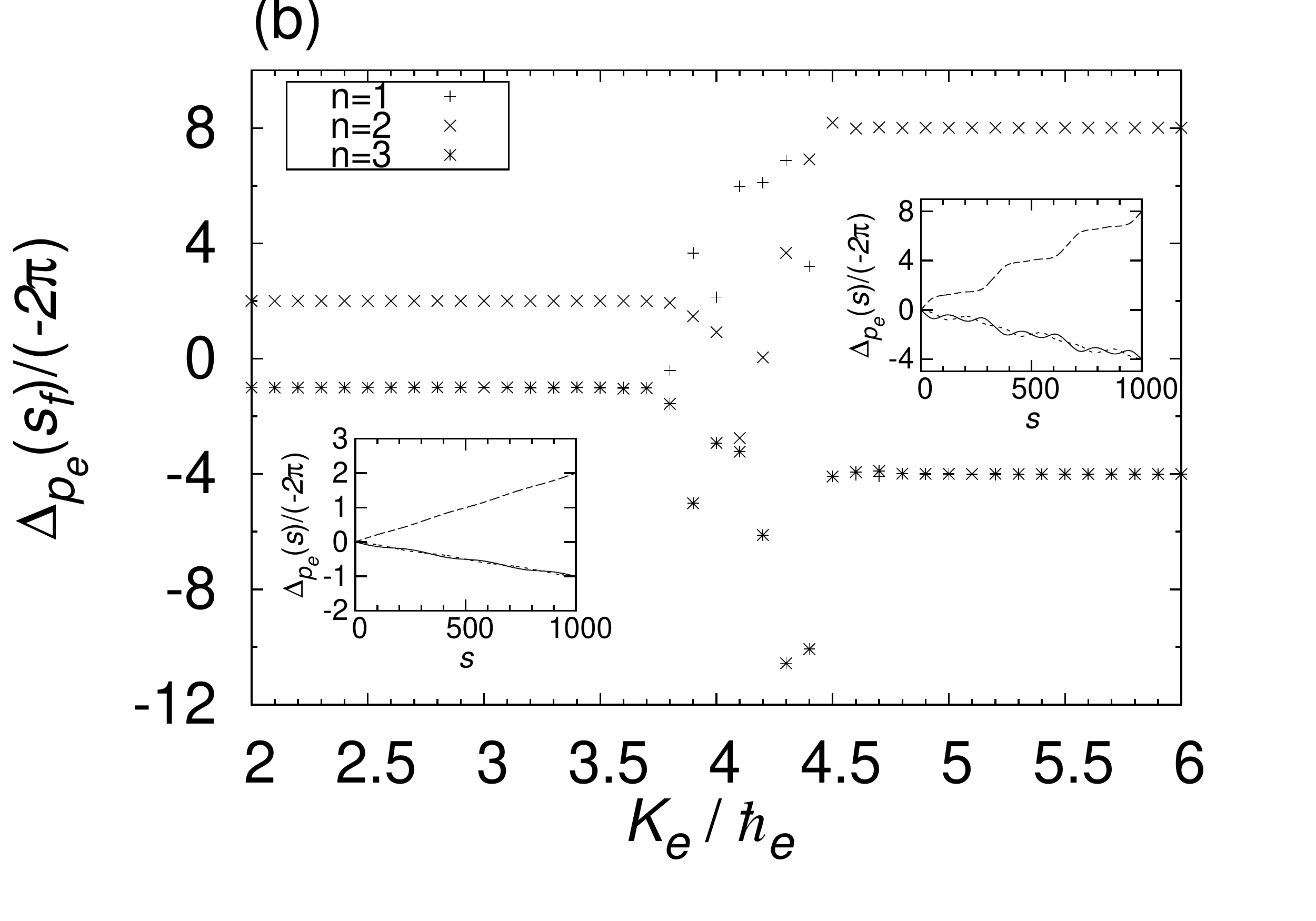} }
} \\
\end{tabular}
\caption{\small{Change in the momentum expectation value (divided by $-2\pi$) vs
$ \eff{K} / \heff $, after one adiabatic cycle implemented in (a) 100  and (b) 1000 discretized steps, for initial
states prepared on each of the three Floquet bands [see Eq.~(\ref{initialstate})].
Insets shows $ \Delta_{\eff{p}}(s)/ (-2\pi)$ vs number of periods $s$,  for $ \eff{K} / \heff = 2.0 $ ($\eff{K} / \heff = 6.0 $) on the left (right).
In each inset, each of the three plotted curves is for one of the three Floquet bands, which in the end
approaches integer values that match the Chern numbers.  }}
\label{transport_numbers}
\end{figure}

 It is necessary to numerically verify our theoretical insights above.  Detailed results are shown in Fig.~\ref{transport_numbers}, again for the case of $\heff=2\pi/3$, with the adiabatic cycle lasting for $s_f=100$ periods (in this case, see also Fig. 3(b) for the final momentum space profile)
 or $s_f=1000$ periods. First of all, apart from the regime of $K_e$ values near the critical point $\eff{K}\approx 4.2 \heff$ (see Fig. 2),
 our numerical values of $ \Delta_{\eff{p}}(s_{f}) /(-2\pi) $ almost perfectly match the Chern numbers. This is the case before or after the jumps
 of the Chern numbers. The insets of both panels depict how $\Delta_{\eff{p}}(s) /(-2\pi)$ builds up with time
 and eventually reaches integer values that match the Chern numbers. In addition, we have checked numerically that if we repeat the adiabatic cycle, then the same quantized increase in momentum expectation value is obtained \cite{supple}.
 We are thus witnessing a clear quantization effect in acceleration as an outcome of Floquet band topology.
 Note however, in the vicinity of phase transition points, e.g., $\eff{K}/\eff{\hbar}\approx 4.2$,
 momentum space transport is no longer quantized. This is because if a topological phase
 transition is about to occur, then the associated band gaps are not large enough to guarantee adiabaticity.  Supporting this understanding, a comparison
 between Fig.~\ref{transport_numbers} (b) and Fig.~\ref{transport_numbers}(a) shows that a longer adiabatic cycle indeed significantly
 narrows down the non-quantization window.  Our numerical data suggests that at least for the 3-band case here, if the driving field strength $K_e$ is
 far away from the phase transition points, then only 50-100 kicking periods (depending on $K_e$)
 will be needed to observe quantized acceleration.  This is experimentally
 motivating, because Floquet state manipulation itself is a topic of much theoretical interest  \cite{Kohn1997AdiabaticLimitPRA,Dranov1998DiscreteAdiabaticJMathPhys,*Tanaka2011DiscreteAdiabaticPhySocJpn}.  The robustness of this quantization against perturbations is also examined in Appendix \cite{supple}. In short, for the 3-band case above, the quantization effect
 can tolerate about 0.5\% uncertainty
 in $\eff{\hbar}$ and a nonzero $\beta$ around 0.01, which should be achievable in light of previous experiments \cite{Ryu2006highorderPRL}.

In conclusion,  based on extended Floquet bands,
we have exposed topological
phase transitions in driven quantum systems and demonstrated
how quantized adiabatic
transport in momentum space may emerge from Floquet band topology.
Numerical results based on a cold-atom-based dynamical model suggest that future
experimental verification of our results is possible in terms of  initial state preparation,
adiabatic cycle implementation, and the robustness of quantization.

We thank Jiao Wang for providing some useful codes during the early stage of this work.
We thank Adam Zaman Chaudhry, Wayne Lawton, and Lakhaphat Lin Aigu for helpful discussions.  J.G. acknowledges funding from
Academic Research Fund Tier I, Ministry of Education, Singapore
(grant No. R-144-000-276-112).

\pagebreak
\section{Appendix A: Derivation of Quantized Momentum Transport}

We use the same notation as introduced in the main text.
The Floquet propagator of a on-resonance double-kicked rotor is given by
\begin{eqnarray} \label{beta_rotor}
\hat{U} _{R} (\alpha) &=& e^{i{\eff{p}}^{2}/2\heff}e^{-i(\eff{K}/\heff)\mathrm{cos}({q})}e^{-i{\eff{p}}^{2}/2\heff} e^{-i(\eff{K}/ \heff)\mathrm{cos}({q}+\alpha)},
\end{eqnarray}
with eigenstates and eigenvalues denoted by
\begin{equation}
\hat{U} _{R} (\alpha) \ket{\psi _{n} (\phi,\alpha)} = e^{i\omega _{n} (\phi,\alpha)} \ket{\psi _{n} (\phi,\alpha)}.
\end{equation}
Here $ \ket{\psi _{n} (\phi,\alpha)} $ represent the Bloch eigenstates
 of $ \hat{U} _{R} (\alpha) $ with Bloch phase $ \phi $, due to momentum-space periodicity of period $N\hbar_e$ for $\hbar_e=2\pi M/N$.
 Using the Bloch theorem,  one rewrites $\langle m  \ket{\psi _{n} (\phi,\alpha)}$
 as \begin{equation} \label{blochform}
\langle m \ket{\psi _{n} (\phi,\alpha)} \equiv e^{i m \phi / N } \langle m \ket{u_{n} (\phi ,\alpha)}.
\end{equation}
Here $ \ip{m+N}{u_{n} (\phi ,\alpha)} = \ip{m}{u_{n} (\phi ,\alpha)}$ holds for all momentum eigenstates $ \ket{m} $. We adopt the normalization convention $ \sum _{m=1} ^{N} |\ip{m}{u_{n} (\phi ,\alpha)}|^{2} = 1 $. With this normalization convention, it can be shown that the identity operator in terms of the Bloch eigenstates is given by
\begin{equation} \label{identity}
\frac{1}{2\pi} \int _{0} ^{2\pi} d\phi \sum _{n=1} ^{N} \ket{\psi _{n} (\phi,\alpha)} \bra{\psi _{n} (\phi,\alpha)} = \mathds{1},
\end{equation}
and their inner products are given by
\begin{equation} \label{innerprod}
\ip{\psi_{n'} (\phi ')}{\psi_{n} (\phi)} = 2 \pi \delta _{n,n'} \delta(\phi-\phi '),
\end{equation}
which implies that the initial states we considered in the main text [see also Eq. (\ref{specialstate}) below] are normalized to unity. Note that we have chosen a basis of Bloch eigenstates, $ \{ \ket{\psi _{n} (\phi,\alpha)} \} $, which are set to be periodic functions of $\phi$, meaning that  $ \ket{\psi _{n} (\phi=0,\alpha)}=\ket{\psi _{n} (\phi=2\pi,\alpha)} $ for all $\alpha$.
Also, the overall phase convention of $ \{ \ket{\psi _{n} (\phi,\alpha)} \} $ can be chosen such that it is a locally single-valued function of $\phi$
and $\alpha$.

Next, we consider the states involved in the adiabatic cycle described in the main text. Such states are of the form
\begin{equation} \label{specialstate}
\ket{\Psi_{n} (\alpha_{s}) } = \frac{1}{2\pi} \int ^{2\pi} _{0} d\phi \ket{\psi _{n} (\phi,\alpha_{s})} e^{i\theta(\phi,\alpha_{s})},
\end{equation}
where $s$ takes values from $0$ to $s_{f}$ as explained in the main text and $ \theta(\phi,\alpha=0) = 0$.
We begin our derivation of quantized transport in momentum space by taking momentum expectation value of such states,
\begin{equation}
\bra{\Psi_{n} (\alpha_{s})} \eff{p} \ket{\Psi _{n} (\alpha_{s})} = \sum _{m = -\infty} ^{\infty} m \heff \ip{m}{\Psi_{n} (\alpha_{s})} \ip{\Psi_{n} (\alpha_{s})}{m}.
\end{equation}
We insert two copies of Eq. (\ref{identity}), one in the middle of each of the inner products of the right hand side. Then, making use of Eqs. (\ref{specialstate}), (\ref{innerprod}) and (\ref{blochform}), we arrive at
 \begin{eqnarray}
\bra{\Psi_{n} (\alpha_{s})} \eff{p} \ket{\Psi _{n} (\alpha_{s})} &=& \sum _{m=-\infty} ^{\infty} m \heff \int _{0} ^{2\pi} d\phi \int _{0} ^{2\pi} d\phi ' e^{\frac{im(\phi-\phi ')}{N}} \ip{m}{u_{n} (\phi,\alpha_{s})} \ip{u_{n}(\phi ',\alpha_{s})}{m} \nonumber \\
&& \times \: \frac{1}{4 \pi^{2}}e^{i(\theta(\phi,\alpha)-\theta(\phi ', \alpha))}.
\end{eqnarray}
We then rewrite $m$ as a derivative in $\phi$ acting on the exponential term and perform an integration by parts to shift this derivative off the exponential and onto the other $\phi$-dependent terms. Further, we split $m$ into $m' + sN$, where $m'$ takes integer values $1,2, \cdots, N$, and $s$ takes all integer values from $-\infty$ to $\infty$. Then, after making use of the fact that $\ip{m+sN}{u_{n} (\phi,\alpha)}=\ip{m}{u_{n} (\phi,\alpha)} $ for all momenta $m \heff $ and all $ s \in \mathds{Z}$, one invokes Poisson's summation formula $ \sum _{s= -\infty} ^{\infty} e^{is(\phi-\phi ')} = \sum _{s= -\infty} ^{\infty} 2\pi \delta (\phi - \phi ' - s \: 2\pi)$ and finds
\begin{eqnarray}
&&\frac{1}{2\pi} \sum_{m'=1} ^{N}  \int _{0} ^{2\pi} d\phi \ip{u_{n} (\phi,\alpha_{s})}{m'} Ni\heff \frac{\partial}{\partial \phi} (\ip{m'}{u_{n} (\phi,\alpha_{s})}) \nonumber \\
&=& \frac{1}{2\pi} \int _{0} ^{2\pi} d\phi \bra{\bar{u}_n(\phi,\alpha_{s})} Ni\heff \frac{\partial}{\partial \phi} \ket{\bar{u}_n(\phi,\alpha_{s})},
\end{eqnarray}
where $\ket{\bar{u}_n(\phi,\alpha)}\equiv
\sum_{m=1}^{N} \ket{m} \ip{m}{u_{n} (\phi ,\alpha)}$ and we have made use of the fact that $\int ^{2\pi} _{0} d \phi  \frac{\partial \theta(\phi,\alpha)}{\partial \phi} = 0 $.
Summarizing, we have shown that
\begin{eqnarray} \label{p_expectatn}
\langle \Psi_{n}(\alpha_{s})| \eff{p} | \Psi_{n}(\alpha_{s})\rangle & =& \frac{1}{2\pi}\int ^{2\pi} _{0} d \phi  \langle \bar{u}_{n} (\phi ,\alpha_{s})| Ni \heff \frac{\partial}{\partial \phi} |\bar{u} _{n} (\phi ,\alpha_{s})\rangle.
\end{eqnarray}
Note that $\ket{\bar{u}_n(\phi,\alpha _{s}}$ are the eigenstates of the reduced Floquet matrix
\begin{equation}
\langle m|\bar{U}(\phi,\alpha_{s})|m'\rangle \equiv \sum^{\infty} _{l = - \infty} \langle m| \hat{X}^{\dagger} (\phi) \hat{U}_{R} (\alpha_{s}) \hat{X}(\phi)| m'+lN\rangle,
\end{equation}
with $m,m' = 1, \cdots, N$ and $\hat{X}(\phi)\equiv e^{ip_{e} \phi/N \hbar_{e}}$.

Next, we consider the difference in momentum expectation value between consecutive states in the adiabatic cycle, namely,
\begin{equation}\label{small_p_change}
\delta \langle \eff{p} \rangle _{s} \equiv \bra{\Psi_{n} (\alpha_{s})} \eff{p} \ket{\Psi _{n} (\alpha_{s})}-\bra{\Psi_{n} (\alpha_{s-1})} \eff{p} \ket{\Psi _{n} (\alpha_{s-1})}.
\end{equation}
For both terms in the above equation, the expectation values are given by Eq.~(\ref{p_expectatn}) with $\alpha=\alpha_{s-1}$ and $\alpha=\alpha_{s}$ respectively.  Next one may express
$ \ket{u_{n} (\phi,\alpha _{s}) } $ in terms of $ \ket{u_{n} (\phi,\alpha _{s-1})}$ to the first order in $\delta \alpha =\alpha_s -\alpha_{s-1}$, i.e.,
\begin{equation} \label{1st_order_state}
 \ket{\bar{u} _{n} (\phi ,\alpha _{s})}
 = \ket{\bar{u} _{n} (\phi ,\alpha _{s-1})} + \ket{\bar{u}_{n} ^{(1)} (\phi,\alpha_{s-1})} \,
\end{equation}
where
\begin{equation}
\ket{\bar{u}_{n} ^{(1)} (\phi,\alpha_{s-1})} \equiv  \delta\alpha \sum_{n'=1,\neq n} ^{N} \frac{\langle \bar{u} _{n'} (\phi ,\alpha _{s-1})| \frac{\partial \bar{U}(\phi,\alpha _{s-1})}{\partial \alpha_{s-1}} | \bar{u} _{n} (\phi ,\alpha _{s-1})\rangle }{e^{i \omega _{n} (\phi,\alpha_{s-1})}-e^{i \omega _{n'} (\phi,\alpha_{s-1})}} \: \ket{\bar{u} _{n'} (\phi ,\alpha _{s-1})}.
\label{correction}
\end{equation}
This perturbation theory can be performed because (i) our Bloch eigenstates can be chosen as continuous in $\alpha$ for any fixed $\phi$ and (ii)
in the adiabatic limit $s_f\rightarrow +\infty$, $\delta\alpha\rightarrow 0$ and hence only first-order matters.
 It is now seen that
 $\ket{\bar{u}_n(\phi,\alpha_{s})}$ is expressed as the eigenstate (correct to the first order in $\delta \alpha$) of the reduced Floquet matrix with $\alpha = \alpha_{s-1}$, plus a correction.  Note also that the expression $\frac{\partial \bar{U}(\phi,\alpha _{s-1})}{\partial \alpha_{s-1}}$ in the correction term
can be explicitly calculated by $\hat{U}_{\alpha}$ via
\begin{equation}
\frac{\partial}{\partial \alpha_{s-1}} \langle m|\bar{U}(\phi,\alpha _{s-1})|m'\rangle  \equiv \sum^{\infty} _{l = - \infty} \langle m| \hat{X}^{\dagger} (\phi) \frac{\partial \hat{U}_{R} (\alpha _{s-1})}{\partial \alpha_{s-1}} \hat{X}(\phi)| m'+lN\rangle .
\end{equation}
%added to it (which is just the first-order approximation of the reduced Floquet matrix with $\alpha = \alpha_{s}$).
Finally, we use the Feynman-Hellmann-like formula
\begin{equation} \label{feynhell_like}
\left( \frac{\partial}{\partial \phi} \bra{\bar{u}_{n} (\phi,\alpha)}\right)\ket{\bar{u}_{n'} (\phi,\alpha)} = \frac{\bra{\bar{u}_{n} (\phi,\alpha)} \frac{\partial \bar{U}^{\dagger} (\phi,\alpha)}{\partial \phi}\ket{\bar{u} _{n'} (\phi,\alpha)}}{e^{-i\omega _{n}}-e^{-i\omega _{n'}}},
\end{equation}
and its adjoint. Substituting Eq.~(\ref{feynhell_like}), its adjoint, and Eq.~(\ref{1st_order_state}) into Eq.~(\ref{small_p_change}) and keeping only first order terms in $\delta \alpha$, we obtain
\begin{eqnarray}
 \delta \langle {\eff{p}} \rangle _{s} = -\frac{1}{2\pi} N \heff \int _{0} ^{2\pi} d\phi \: B _{n} (\phi,\alpha_s) \: \delta \alpha ,
\end{eqnarray}
where $ B _{n} (\phi,\alpha)$ is identified as the Berry curvature
\begin{eqnarray}
 B _{n} (\phi,\alpha) =  i \sum _{n'=1,\neq n} ^{N} \Bigg\{ \frac{\langle \bar{u}_{n}|\frac{\partial \bar{U}^{\dagger}}{\partial  \phi}|\bar{u}_{n'}\rangle\langle \bar{u}_{n'}
|\frac{\partial \bar{U}}{\partial \alpha}|\bar{u}_{n}\rangle}{|e^{i \omega _{n}}-e^{i \omega _{n'}}|^{2}} - \mathrm{c.c} \Bigg\},
\end{eqnarray}
with the explicit dependence of $|\bar{u}_{n}(\phi ,\alpha)\rangle$, $\omega _{n} (\phi,\alpha)$ and $\bar{U}(\phi,\alpha)$ on $\alpha$ and $\phi$ all
suppressed for brevity.

The total change in the expectation value of ${p}_e$ from $t=0$ to $(sT)^{-}$ is denoted by $ \Delta_{p_e}(s)$.
Because $ \Delta_{p_e}(s)= \sum _{s'=1} ^{s} \delta \langle {\eff{p}} \rangle _{s'} $,
one finds the total momentum change over the entire adiabatic cycle, i.e.,
\begin{eqnarray}
\Delta_{p_e}(s_{f})  &=& -  M \int _{0}^{2\pi} d\phi \int _{0}^{2\pi} d\alpha \: B _{n} (\phi,\alpha) \nonumber \\
&= &- 2 \pi M C_{n},
\label{finaleq}
\end{eqnarray}
where we have used $\heff = 2\pi M/N$ and $ C _{n} $ is exactly the Chern number of the $n$th Floquet band defined earlier, whose equivalent
expression is
\begin{eqnarray}
C_{n} = \frac{1}{2 \pi} \int _{0} ^{2\pi} d\phi \int _{0} ^{2\pi} d\alpha \: B _{n} (\phi,\alpha).
 \end{eqnarray}
 This completes our derivation of quantization of transport in momentum space.

 \section{Appendix B: Details of Chern Number Calculation}

Here we present some details regarding how the Chern numbers of Fig. 2 of the main text are calculated. In terms of a line integral, the Chern number is given by
\begin{equation}
C_{n} = \frac{i}{2\pi} \oint  \bra{\bar{\psi} _{n} ( \Gamma )} \frac{\partial}{\partial \Gamma} \ket{\bar{\psi}_{n} (\Gamma) } d \Gamma, \label{chern_num_defn}
\end{equation}
where $ \Gamma \equiv (\phi, \alpha) $, $\ket{\bar{\psi}_{n} (\Gamma) } \equiv \sum _{m=1} ^{N} \ket{m}\ip{m}{\psi _{n} (\Gamma)}$ (normalized on $N$ momentum
components).
We first mention that this expression is equivalent to
$\frac{i}{2\pi} \oint  \bra{\bar{u} _{n} ( \Gamma )} \frac{\partial}{\partial \Gamma} \ket{\bar{u}_{n} (\Gamma) } d \Gamma $, where $|\bar{u}_n(\Gamma)\rangle\equiv
\sum_{m=1}^{N} |m\rangle\langle m \ket{u_{n} (\Gamma)}$, and $\ket{\bar{\psi}_{n} (\Gamma) } =  \hat{X}(\phi) \ket{\bar{u}_{n} (\phi ,\alpha)}$, with $ \hat{X} (\phi) \equiv e^{i\eff{p}\phi / N\heff }$.  To see this, one needs only to substitute $\ket{\bar{\psi}_{n} (\Gamma) } =  \hat{X}(\phi) \ket{\bar{u}_{n} (\phi ,\alpha)}$ into the Chern number expression above and write the integral as the sum of 4 terms, each on one edge of the $(\phi,\alpha)$ Brillouin zone. The two terms containing $\frac{\partial}{\partial \alpha}$ immediately reduce to the form $\frac{i}{2\pi} \int  \bra{\bar{u} _{n} ( \Gamma )} \frac{\partial}{\partial \alpha} \ket{\bar{u}_{n} (\Gamma) } d \alpha $ because $\hat{X} (\phi)$ is not a function of $\alpha$. The other two terms containing $\frac{\partial}{\partial \phi}$ will also reduce to the similar form $\frac{i}{2\pi} \int  \bra{\bar{u} _{n} ( \Gamma )} \frac{\partial}{\partial \phi} \ket{\bar{u}_{n} (\Gamma) } d \phi $ after one application of the product rule of differentiation and noting that two of the resulting four terms cancel. This then leaves us with  $ C_{n} =\frac{i}{2\pi} \oint  \bra{\bar{u} _{n} ( \Gamma )} \frac{\partial}{\partial \Gamma} \ket{\bar{u}_{n} (\Gamma) } d \Gamma $.  We can now work on $C_{n}$ in terms of $\ket{\bar{u}_{n}}$ to make the link with the main text more concrete, though it should be clear that it is entirely valid to work with $\ket{\bar{\psi}_{n}}$ instead. In the remainder of this section, we will derive the surface integral formula for the Chern number as it was used in the main text and then mention some details on our numerical calculation.

The surface integral for $C_{n}$ which involves the Berry curvature as shown in the main text is obtained from the line integral via the steps shown here. First, a single application of Stoke's theorem leaves us with
\begin{eqnarray}
C_{n}&=&\frac{i}{2\pi} \int _{0} ^{2\pi} \int _{0} ^{2\pi} d\phi d\alpha \left[ \frac{\partial}{\partial \phi} \left(\bra{\bar{u}_{n}(\phi,\alpha)} \frac{\partial}{\partial \alpha}\ket{\bar{u}_{n}(\phi,\alpha)}\right)-\frac{\partial}{\partial \alpha} \left(\bra{\bar{u}_{n}(\phi,\alpha)} \frac{\partial}{\partial \phi}\ket{\bar{u}_{n}(\phi,\alpha)} \right) \right] \nonumber \\
&=& \frac{i}{2\pi} \int _{0} ^{2\pi} \int _{0} ^{2\pi} d\phi d\alpha \left[ \left( \frac{\partial}{\partial \phi}\bra{\bar{u}_{n}(\phi,\alpha)} \right) \frac{\partial}{\partial \alpha}\ket{\bar{u}_{n}(\phi,\alpha)} -\left(\frac{\partial}{\partial \alpha}\bra{\bar{u}_{n}(\phi,\alpha)}\right) \frac{\partial}{\partial \phi}\ket{\bar{u}_{n}(\phi,\alpha)}\right]. \nonumber \\
\label{tedious_formula}
\end{eqnarray}
We then insert one copy of the $N \times N$ identity matrix of the form $\sum _{n'} \ket{\bar{u}_{n'}}\bra{\bar{u}_{n'}} = \mathds{1} _{N\times N}$ between each of the two inner products, where we have refrained from writing the functional dependence of $\ket{\bar{u}_{n}}$ on $(\phi,\alpha)$ for brevity. The term with $n'=n$ in one inner product will cancel that in the other inner product because $ \left(\frac{\partial}{\partial \phi}\bra{\bar{u}_{n}}\right) \ket{\bar{u}_{n}} = -\bra{\bar{u}_{n}} \left(\frac{\partial}{\partial \phi}\ket{\bar{u}_{n}}\right) $ and similarly for $\frac{\partial}{\partial \alpha} $. This leads to
\begin{equation}
C_{n}=\frac{i}{2\pi} \int _{0} ^{2\pi} \int _{0} ^{2\pi} d\phi d\alpha \sum_{n' \neq n} ^{N} \left[ \left( \frac{\partial}{\partial \phi}\bra{\bar{u}_{n}} \right) \ket{\bar{u}_{n'}}\bra{\bar{u}_{n'}}\frac{\partial}{\partial \alpha}\ket{\bar{u}_{n}} -\left(\frac{\partial}{\partial \alpha}\bra{\bar{u}_{n}(\phi,\alpha)}\right)\ket{\bar{u}_{n'}}\bra{\bar{u}_{n'}}\frac{\partial}{\partial \phi}\ket{\bar{u}_{n}}\right].
\end{equation}
To proceed, we note that the states $\ket{\bar{u}_{n}(\phi,\alpha)}$ are eigenstates of the reduced Floquet matrix
\begin{equation}\label{reduced_Floquet}
\langle m|\bar{U}(\phi,\alpha)|m'\rangle \equiv \sum^{\infty} _{l = - \infty} \langle m| \hat{X}^{\dagger} (\phi) \hat{U}_{R} (\alpha) \hat{X}(\phi)| m'+lN\rangle,
\end{equation}
with $m,m' = 1, \cdots, N$, and where $\hat{U}_{R} (\alpha)$ is the ``on-resonance double kicked rotor model'' Floquet operator defined in the main text.
From the eigenvalue equation $ \bar{U} (\phi,\alpha)\ket{\bar{u}_{n}(\phi,\alpha)} =e^{i\omega_{n}(\phi,\alpha)}\ket{\bar{u}_{n}(\phi,\alpha)}$, we may take a derivative with respect to $\phi$ on both sides, take the inner product with $\bra{\bar{u}_{n'\neq n} (\phi,\alpha)}$ from the left on both sides, and rearrange to obtain the Feynman-Hellmann-like formula
\begin{equation}
\bra{\bar{u}_{n'} (\phi,\alpha)} \frac{\partial}{\partial \phi} \ket{\bar{u}_{n} (\phi,\alpha)} = \frac{\bra{\bar{u}_{n'} (\phi,\alpha)}\frac{\partial \bar{U} (\phi,\alpha)}{\partial \phi}\ket{\bar{u}_{n} (\phi,\alpha)}}{e^{i\omega_{n}}-e^{i\omega_{n'}}}, \label{FH}
\end{equation}
which has adjoint given by
\begin{equation}
\left( \frac{\partial}{\partial \phi} \bra{\bar{u}_{n} (\phi,\alpha)}\right)\ket{\bar{u}_{n'} (\phi,\alpha)} = \frac{\bra{\bar{u}_{n} (\phi,\alpha)} \frac{\partial \bar{U}^{\dagger} (\phi,\alpha)}{\partial \phi}\ket{\bar{u} _{n'} (\phi,\alpha)}}{e^{-i\omega _{n}}-e^{-i\omega _{n'}}}. \label{FH_adjoint}
\end{equation}
Similar expressions also hold for the $\frac{\partial}{\partial \alpha}$ derivative. We substitute the Feynman-Hellman-like formulas and their adjoints for both $\phi$ and $\alpha$ into Eq. (\ref{tedious_formula}) and obtain
\begin{equation}
C_{n} = \frac{1}{2\pi} \int _{0} ^{2\pi} \int _{0} ^{2\pi} d\phi d\alpha \: B_{n} (\phi,\alpha), \label{chern_surface_integral}
\end{equation}
where
\begin{equation}
B _{n} (\phi,\alpha) =  i \sum _{n'=1,\neq n} ^{N} \Bigg\{ \frac{\langle \bar{u}_{n}|\frac{\partial \bar{U}^{\dagger}}{\partial  \phi}|\bar{u}_{n'}\rangle\langle \bar{u}_{n'}
|\frac{\partial \bar{U}}{\partial \alpha}|\bar{u}_{n}\rangle}{|e^{i \omega _{n}}-e^{i \omega _{n'}}|^{2}} - \mathrm{c.c} \Bigg\}
\end{equation}
is the Berry curvature. This completes the derivation of the surface integral for $C_{n}$ featured in the main text.

We now mention some brief details on how we numerically calculate the Chern numbers. We chose Eq. (\ref{chern_surface_integral}) to calculate the Chern numbers rather than Eq. (\ref{chern_num_defn}) because the former method allows us to use any phase convention for the $\ket{\bar{u}_{n}}$ eigenstates (a quick check shows that the Berry curvature is invariant with respect to replacing each eigenstate with itself multiplied by a global phase). First, we discretize the $(\phi,\alpha)$ Brillouin zone into a sufficiently fine grid. We then numerically work out the integrand of Eq. (\ref{chern_surface_integral}) at each point on the grid and numerically integrate to obtain the Chern number. The different parts of the integrand are obtained as follows. We numerically evaluate the $N \times N$ elements of the $\bar{U} (\phi,\alpha)$ matrix via Eq. (\ref{reduced_Floquet}). The infinite sum on the right hand side there does not present a problem because $ \langle m | \hat{U} _{R} (\alpha) | n \rangle $ decays rapidly as $|m-n|$ increases, so we can truncate the sum at some suitable point with practically no loss of accuracy. Each term $ \langle m | \hat{U} _{R} (\alpha) | n \rangle $ is efficiently calculated via Fast Fourier Transform (FFT). We then numerically diagonalize and obtain the $N$-element vectors representing the $\ket{\bar{u}_{n}}$ eigenstates as well as the eigenphases $\omega_{n}$. Lastly, we obtain the elements of the $ N \times N $ matrices $ \frac{\partial \bar{U}}{\partial \phi}$ and $\frac{\partial \bar{U}}{\partial \alpha}$. This is done by acting $\frac{\partial}{\partial \phi}$ and $\frac{\partial}{\partial \alpha}$ on both sides of Eq. (\ref{reduced_Floquet}), simplifying the expression on the right hand side by hand, and then obtaining the numerical values via FFT. The infinite sums here may again be truncated at a suitable level with no appreciable loss in accuracy.

\section{Appendix C: Stability of Quantized Transport to Perturbations}

To motivate potential experimental interest, in this section we consider the effects of of nonzero quasimomentum variable $\beta$, small perturbations to effective Planck constant $\heff$, and inevitable imperfections in actual experimental implementations  on the quantization of transport in momentum space.

To consider the effects of nonzero $\beta$, we begin from the double-kicked rotor model Floquet propagator,
\begin{eqnarray}
\hat{U} (\alpha) = e ^{-i(T-T_{0})({p} ^{2} / 2 \hbar)}e^{-i(K / \hbar)\mathrm{cos}({q})}e^{-iT_{0}({p} ^{2} / 2 \hbar)}  e^{-i (K / \hbar)\mathrm{cos}({q}+\alpha)},
\label{eq1}
\end{eqnarray}
which describes a kicked particle moving on a line. As explained in \cite{Fishman2003QAM}, the periodicity of the Floquet propagator in $q$ allows us to map the dynamics of a kicked particle on a line onto the dynamics of a fictitious ensemble of kicked particles each on a circle, with every such particle representing the dynamics for a different $\beta$-component of the kicked particle on the line. Each such particle on the circle is thus referred to as a $\beta$-rotor. The $\beta$ rotors may then be evolved separately from one another and recombined appropriately at the end to recover the time-evolved state of the actual particle on the line. The different $\beta$-rotors evolve independently of one another under the action of a $\beta$ Floquet propagator
\begin{eqnarray}
\hat{U}_{\beta} (\alpha) = e ^{-i(T-T_{0})((\hat{n}+\beta\hbar) ^{2} / 2 \hbar)}e^{-i(K / \hbar)\mathrm{cos}({q})}e^{-iT_{0}((\hat{n}+\beta\hbar) ^{2} / 2 \hbar)} e^{-i (K / \hbar)\mathrm{cos}({q}+\alpha)},
\end{eqnarray}
where $\hat{n} + \beta\hbar$ is the $\beta$-rotor's momentum operator with  eigenstates $\ket{m}$ and momentum eigenvalues $(n+\beta)\hbar$ with $n \in \mathds{Z}$. Now, imposing the main quantum resonance condition $T \hbar = 4\pi$, and writing $\heff \equiv \hbar T_{0}$, $\eff{K} \equiv K T_{0}$ and $ \hat{n} _{e} \equiv \hat{n} T_{0}$, the $\beta$ Floquet propagator becomes
\begin{eqnarray}
\hat{U}_{\beta} (\alpha) = e ^{-i(\frac{4\pi}{\heff}-1)[(\hat{n}_{e}+\beta\heff )^{2} / 2 \heff]}e^{-i(\eff{K} / \heff)\mathrm{cos}({q})}e^{-i[(\hat{n}_{e}+\beta\heff) ^{2} / 2 \heff]}  e^{-i (\eff{K} / \heff)\mathrm{cos}({q}+\alpha)}.
\label{betapropagator}
\end{eqnarray}
The rescaled momentum eigenvalues are then given by $(n + \beta)\heff$, $n \in \mathds{Z}$. If we set $\beta = 0$, the momentum operator has eigenvalues of only integer multiples of $\heff$ and we recover the on-resonance double-kicked-rotor-model Floquet propagator described in Eq.~(1) of the main text. For nonzero $\beta$, $ \hat{U}_{\beta}$ in Eq. (\ref{betapropagator}) lacks translational invariance in momentum, so the entire derivation in the previous section will not apply and we hence do not expect to see quantized momentum transport in the $\beta$-rotor dynamics. However, for small $\beta$ values, it is reasonable to expect that evolution of the special initial state in Eq. (\ref{specialstate}) will still show quantized transport up to a good approximation. We find that this is indeed the case, as evidenced in Fig. 5(a), where we see that for $\beta$ values up to $0.01$, the quantized transport still survives for $\eff{K}=2.0\heff$, $\heff=2\pi/3$. Experimentally, however, the initial states prepared are typically a mixture of states with $\beta$ values following a narrow Gaussian distribution peaked at $\beta=0$. Hence, to gauge the experimental feasibility of our proposal, we computed the transport values for a range of $\beta$ values near $0$ and took the average of these values weighted with a Gaussian peaked at $0$. We found that for a Gaussian distribution with standard deviation of $0.01$, the change in momentum expectation value divided by $-2\pi$ for the 3 bands at $\eff{K}=2\heff$ for a 100-period adiabatic cycle are $-0.99$, $1.98$ and $-0.99$ for bands 1,2 and 3, respectively. In other words, the quantization of transport is still observable even with a realistic spread over $\beta$ values.
This indicates that the small $\beta$ spread in current cold-atom experiments based on Bose-Einstein condensates
should not hinder the observation of quantized transport in momentum space.

Next, we briefly consider the effects of small deviations in the Planck constant $\heff$ for the same 3-band case considered above with $\heff=2\pi /3$ and $\eff{K} = 2\heff$.  Certainly, even a slight change in $\heff$ should cause the
Floquet-band structure to change entirely (recall that the Floquet band structure depends on
 whether $\heff$ is rational or irrational), but physical observables should not be as sensitive as the band structure itself.
Indeed, as seen in Fig. 5(b), the quantized transport is still observable for deviations in $\heff$ up to $0.005\heff$.

\begin{figure}[h!]
\begin{center}$
\begin{array}{cc}
%\centering
\includegraphics[width=80mm, height=!]{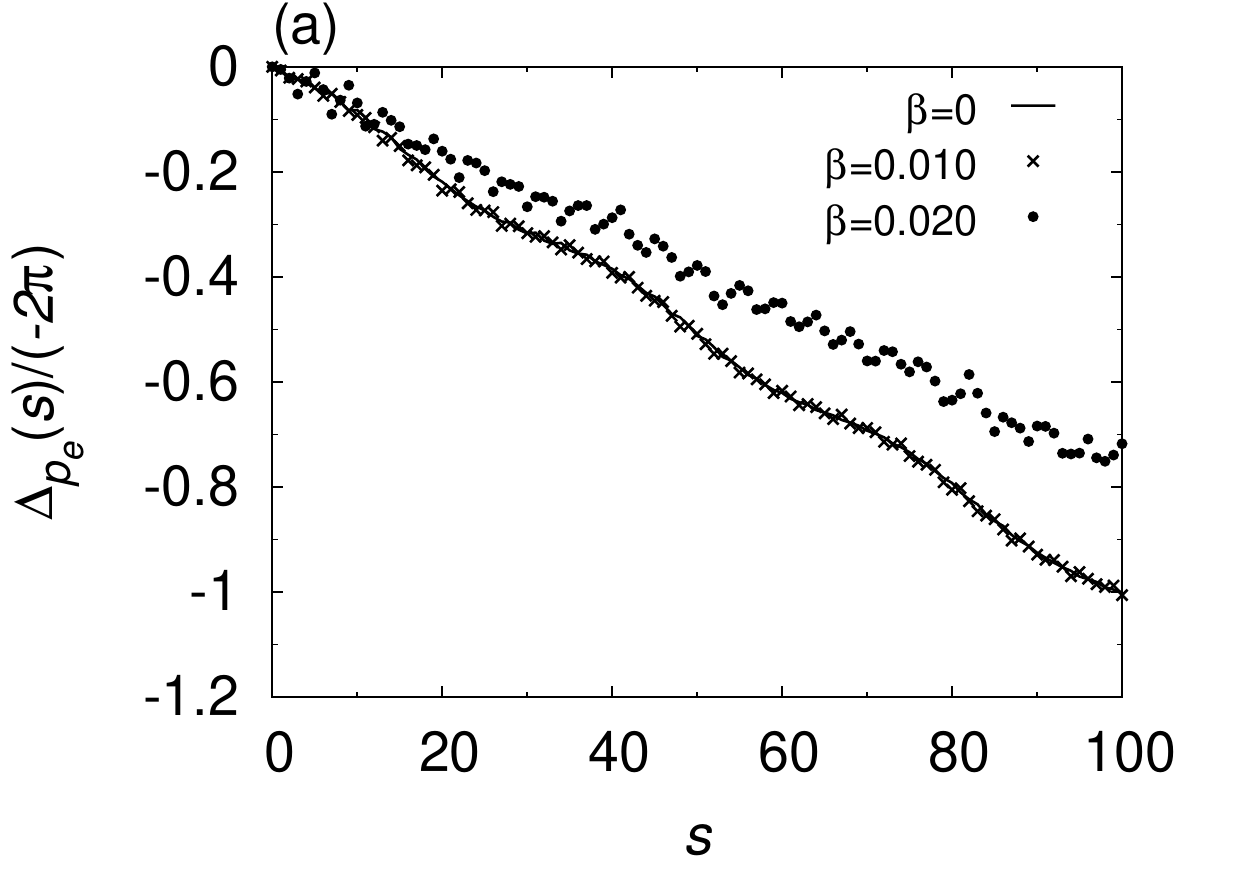}&
\includegraphics[width=80mm, height=!]{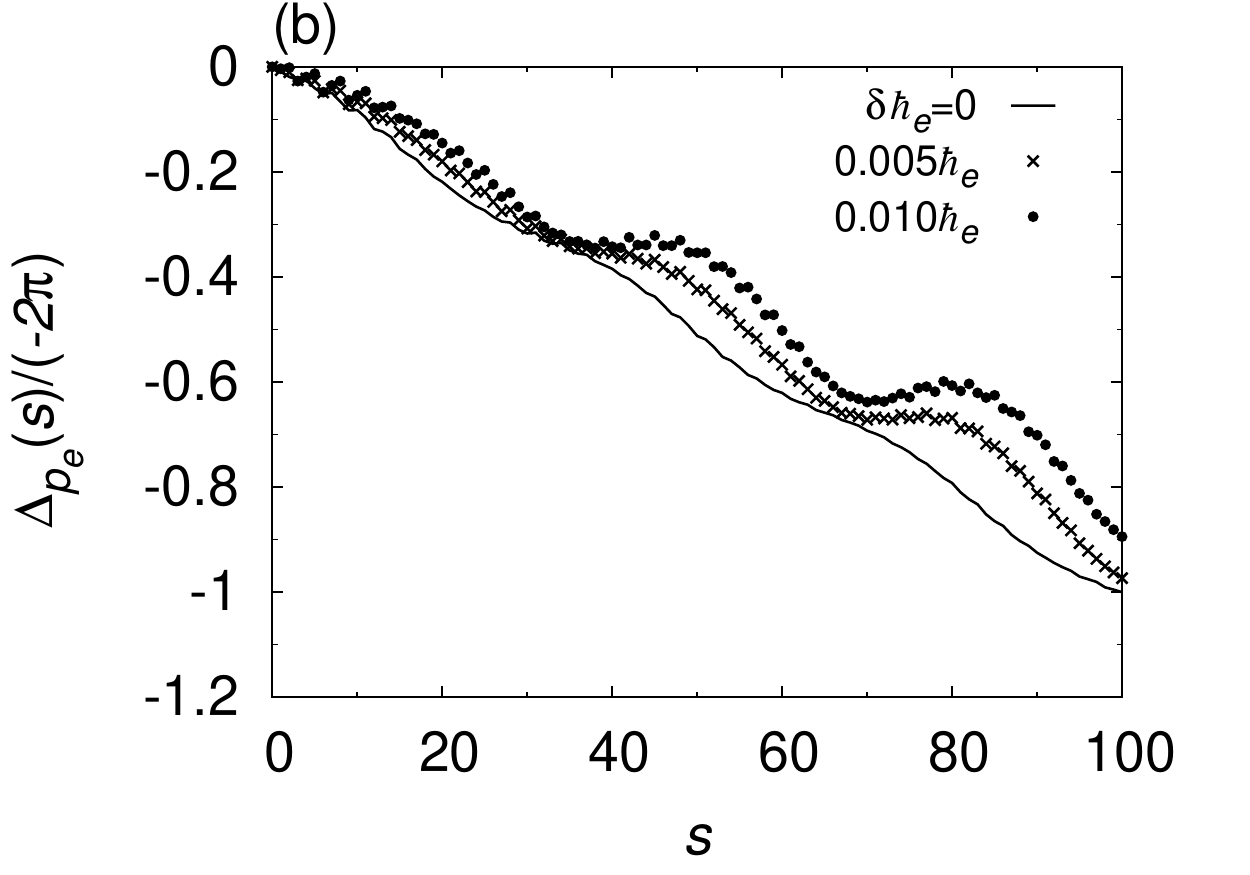}\\
\end{array}$
\end{center}
\caption{(a) Adiabatic momentum transport as a function of time for $\beta= 0, 0.01, 0.02$. (b) Adiabatic momentum transport as a function of time for $\delta \heff = 0, 0.005\heff, 0.01\heff$. In both cases, we consider adiabatic cycles lasting for 100 kicking periods, with the initial state $\ket{\Psi _{3} (\alpha=0)}$ prepared on band 3, for $\eff{K}=2.0\heff$ and $\heff=2\pi/3$. Quantized adiabatic transport is observed in
those cases where the final values of $\Delta_{p_e}$ are close to the integer value $-1$.
} \label{stabilityplots}
\end{figure}

\begin{figure}[h!]
\begin{center}$
\begin{array}{cc}
%\centering
\includegraphics[width=80mm, height=!]{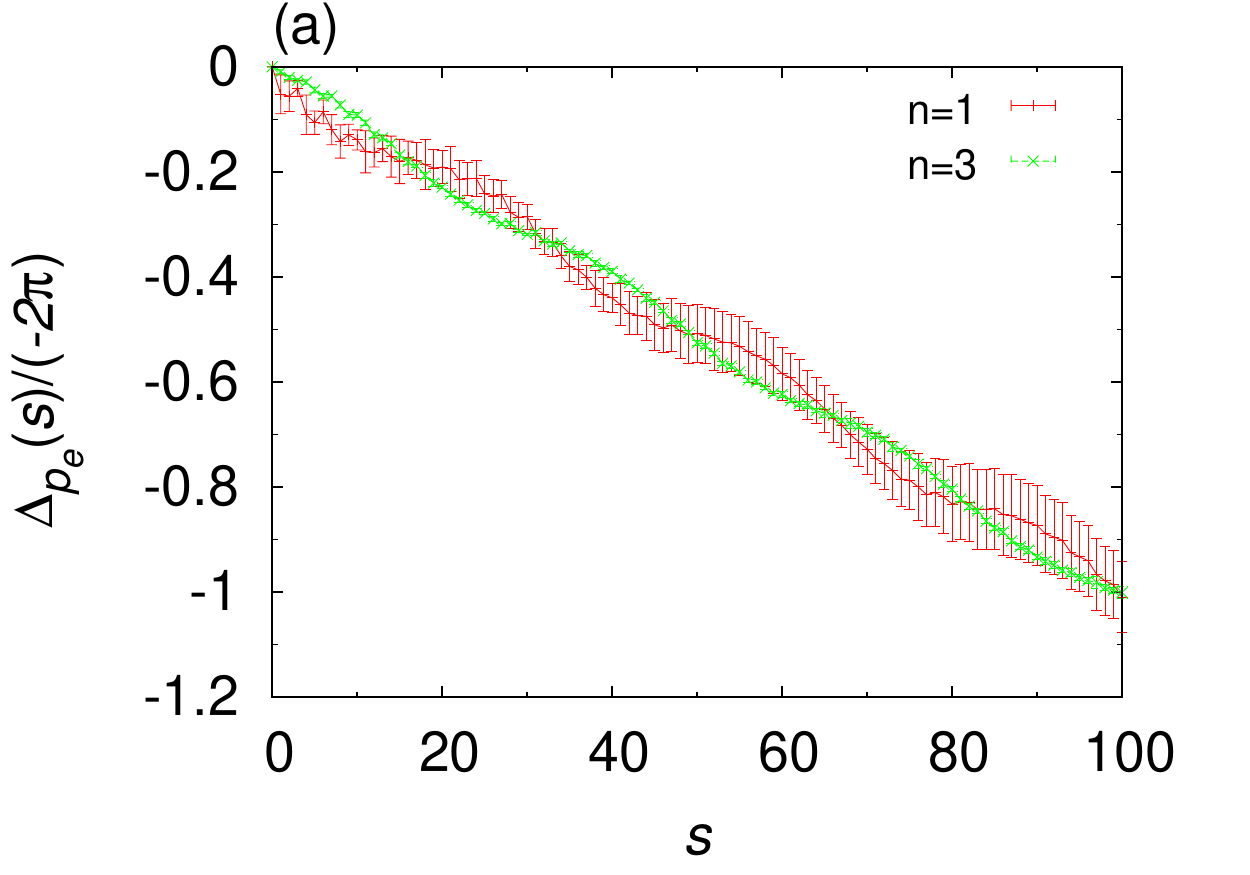}&
\includegraphics[width=80mm, height=!]{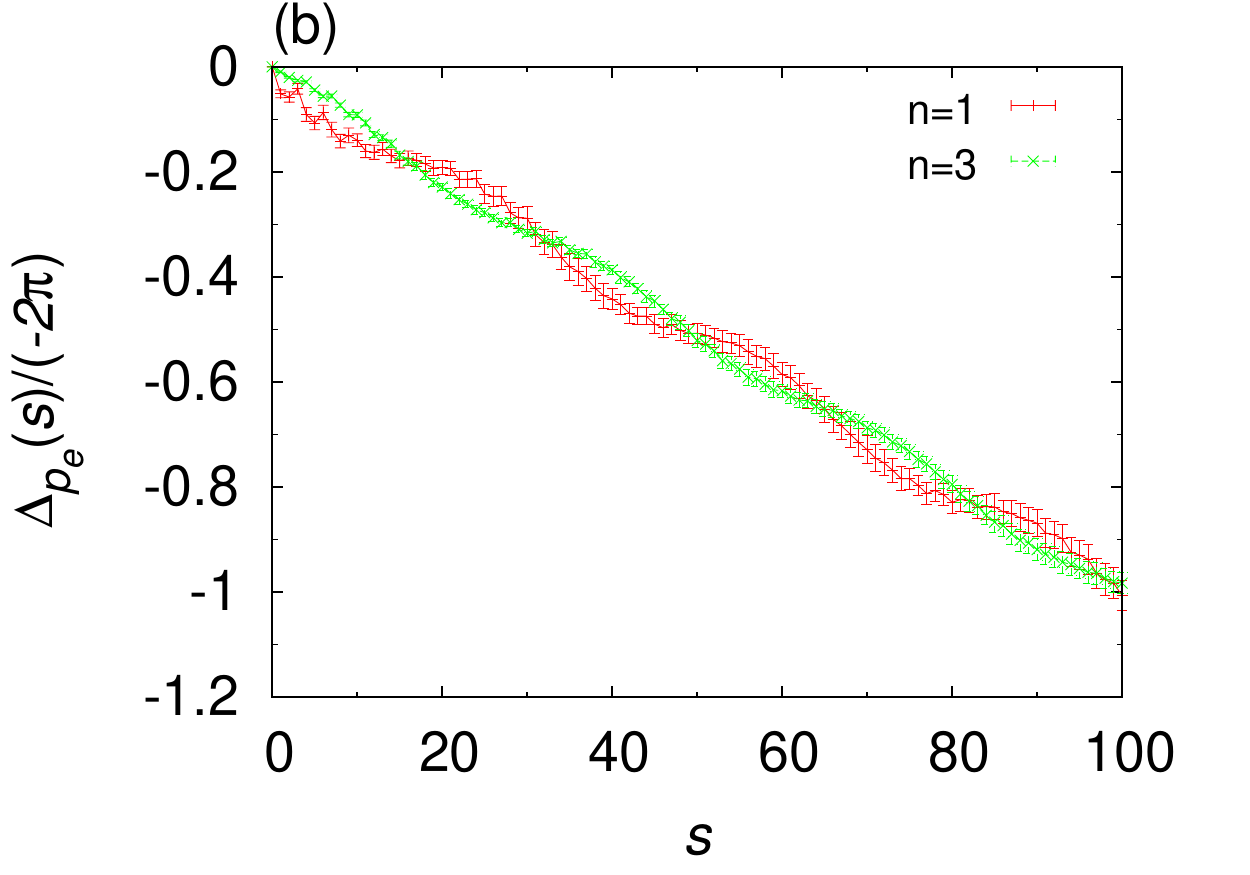}\\
\end{array}$
\end{center}
\caption{Momentum transport as a function of time in the presence of noise (see the text for the noise details and system parameters). The case of panel
 (a) includes noise present in the amplitudes and phases of each momentum eigenstate component in the initial state. The case of panel
  (b) simulates noise present in the value of phase shift $\alpha$ during each step of one adiabatic cycle. In each case, we averaged over
  1000 realizations of the noise. A and B represent noise intensity. The plotted error bars represent the standard deviation in the total
  momentum transport found in our numerical experiment.  Note that the averaged total momentum transport stays very close to the quantized value
  despite the relatively strong noise.
} \label{noiseplots}
\end{figure}

%new segment
Finally, we consider the effects of practical imperfections in experiments. To take into account the difficulty of superposing a large number of different momentum eigenstates in actual experiments, we consider a perturbed version of the initial state $\ket{\Psi _{3} (\alpha=0)}$ by setting to zero all probability amplitudes outside the range $ m \in [-5,5] $ and multiplying the resulting state by an appropriate constant for normalization. This does not alter the original state significantly because its probability is tightly concentrated about $m=0$ (more than $99.9$ \% of the original state's probability lies within $ m \in [-5,5] $) .
With this truncated state as our starting point,  we consider two noise models. In the first model,
we introduce an uncertainty in the relative phases of the 11 surviving momentum states in the superposition by multiplying each one by exp$[i(0.05)2\pi \xi ]$ where $\xi$ is a random variable uniformly distributed in [0,1). We then assume that the probability amplitude of each constituent momentum state is scaled by the term [$1+A(\xi-0.5)$] and then multiply all states by an appropriate constant for normalization.
In our second noise model, we assume that in the $s$-th period, the value of the phase shift $\alpha$ is given by $[s+B(\xi-0.5)]2\pi/100$. We now
present in Fig. 6 two simple numerical examples for band 1 and band 3, with $s_f=100$, $K_e/\hbar_e=2.0$, and $\hbar_e=2\pi/3$.  It is seen that
in both models, the total momentum transport over one cycle (divided by $-2\pi$), when averaged over the 1000 realizations of the random noise,
changes only slightly from the quantized value ($-1$) despite the rather strong noise.
The fair robustness to imperfection to initial state shown in Fig. 6(a) is partially due to the fact
that our initial state is dominated by very few momentum components. Indeed, the fluctuations in the total momentum transport from realization to
 realization, as manifested by the plotted error bars for band 3, are much smaller than
those for band 1, consistent with the fact that our initial state for band 3 turns out to be more localized than that for band 1.
Results in Fig. 6(b) illustrate that slow fluctuations in $\alpha$ (noise
amplitude rather comparable to $2\pi/s_f$) during the adiabatic process do not severely affect the quantization. As seen from
Fig. 6(b), the total momentum transport averaged over 1000 noise histories is only slightly shifted from the quantized value ($-1$), and
fluctuations represented by the error bars are also rather small.  This confirms
that the fine details of an adiabatic process are largely irrelevant.

%Might need another discussion here about actual experiments where all the different types of noise and imperfections all add up.

%end of new segment

\section{Appendix D: The Seven Band Case}

\begin{figure}[h!]
\begin{center}
\includegraphics[trim=1cm 7cm 1cm 5cm,clip=true,height=7.2cm ,width=14cm]{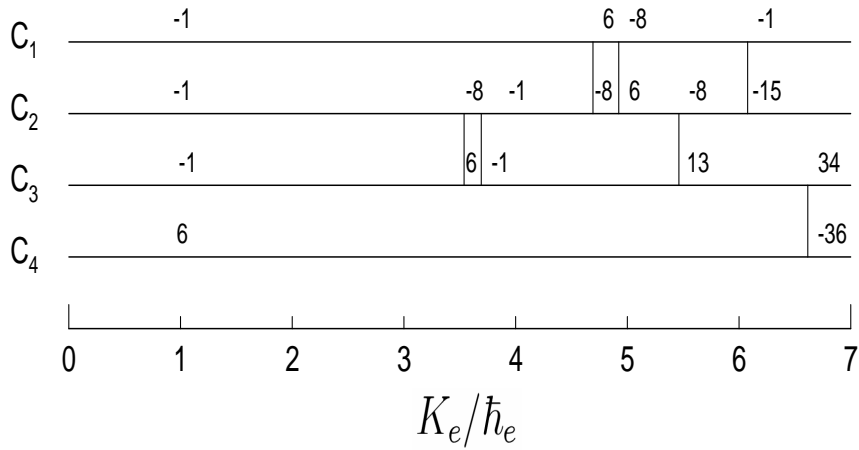}
\end{center}
\caption{The transitions of Chern numbers as kicking strength $K_e$ increases, for $\hbar_e=2\pi/7$.
 The horizontal lines represent the bands and the numbers written on them are their Chern numbers. Vertical lines connecting two horizontal lines represent degeneracies occurring between the two associated bands. Note the unusually big changes of Chern numbers across some critical points.} \label{cherntable}
\end{figure}

In this section, we wish to emphasize that our topological characterization of the Floquet bands using Chern numbers
is equally applicable to cases with more bands.  Here we show a case with 7 bands, with $\heff = 2\pi/7$. We present in Fig. 7 the Chern number results as the kicking strength increases. The associated Chern numbers are found to be symmetric about the central band, so we need only display 4 bands. The same qualitative features as we observed in the 3-band case are seen here. In particular, all bands other than the central band have Chern number $-1$ for low kicking strength, with the central band having a Chern number equal to the negative of the sum of the Chern numbers of all other bands. At band collisions, the Chern number of each band jumps by a multiple of $N=7$ for $\heff= 2\pi /N$.  For this reason we see that the Chern numbers can rapidly become very large integers, which is a quite interesting feature for our system.  We have also carried out adiabatic transport studies for this case. Due to the smaller band gaps for the 7-band case, we need to have longer $\alpha$-cycles (i.e., $\alpha$ must be varied from $0$ to $2\pi$ over a larger number of periods) in order for effective adiabatic following to take place. Experimentally speaking, this case is hence of less interest.  However, in our numerical studies, other than the longer adiabatic cycles, the results
are qualitatively the same as the 3-band case presented in the main text.

\section{Appendix E: Further Numerical Data}

\begin{figure}[h!]

\includegraphics{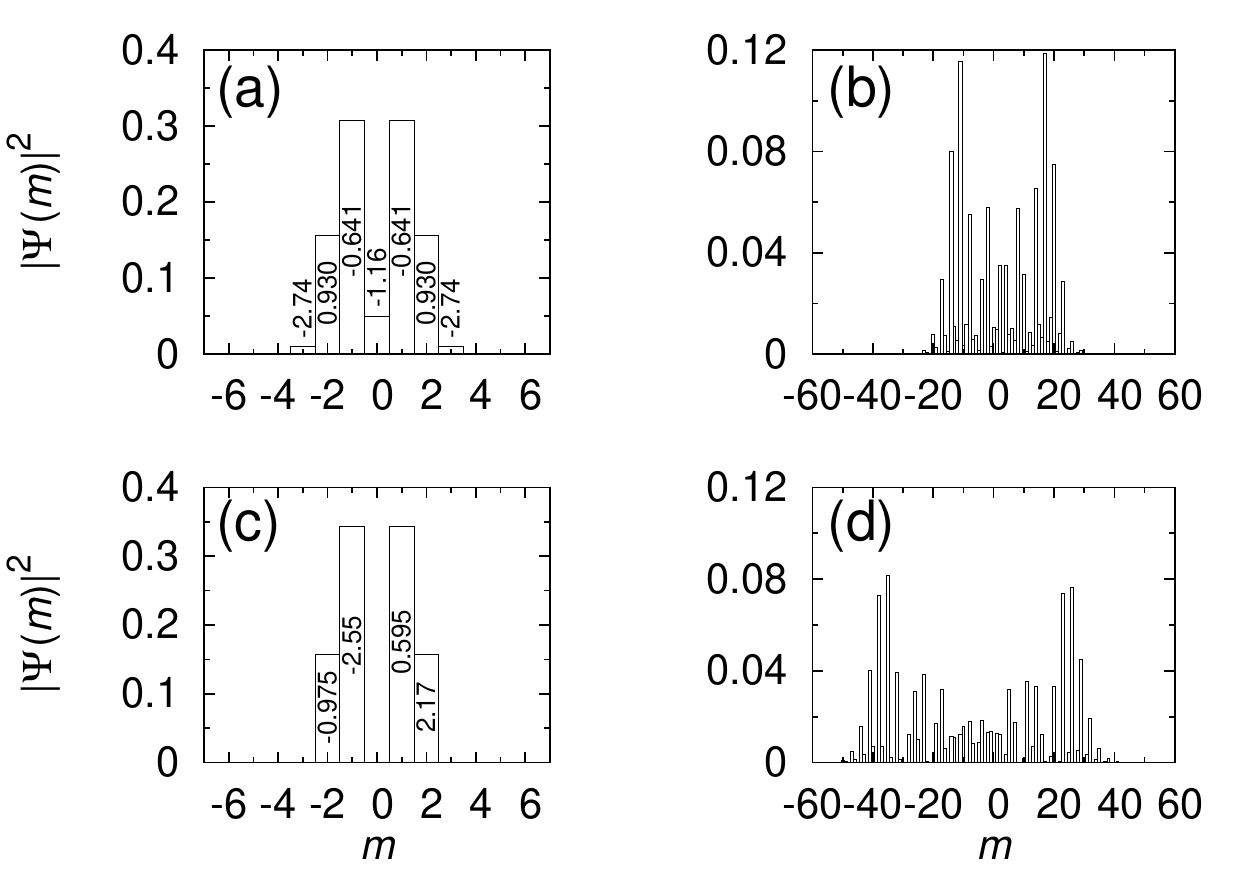}

\caption{\small{For the initial state $ \ket{\Psi_{1} (\alpha = 0)}$ for $ \eff{K} = 2 \heff $ prepared on band 1, the distribution in momentum space before and after a 100-period adiabatic cycle are shown in (a) and (b) respectively. Panels (c) and (d) show the parallel results starting from the initial state $ \ket{\Psi_{2} (\alpha = 0)}$ for $ \eff{K} = 2 \heff $ prepared on band 2. In panels (a) and (c), the numbers displayed are the phases of the amplitudes of the constituent momentum eigenstates in the initial superposition state.}}

\end{figure}

In this last section, we provide some further examples of distributions in momentum space after the adiabatic cycles proposed in the main text. We choose $K_{e} = 2 \hbar _{e} $ for $ \hbar_{e} = 2\pi / 3$, as we did in Fig. 3 of the main text.
Figure 8 shows the initial and final distributions after a single adiabatic cycle for initial Wannier states prepared on bands 1 and 2. Figure 9 shows the distribution in momentum space after completion of a second and third adiabatic cycle for an initial Wannier state prepared on band 3 (see Fig. 3 of the main text).

\begin{figure}[h!]

\includegraphics{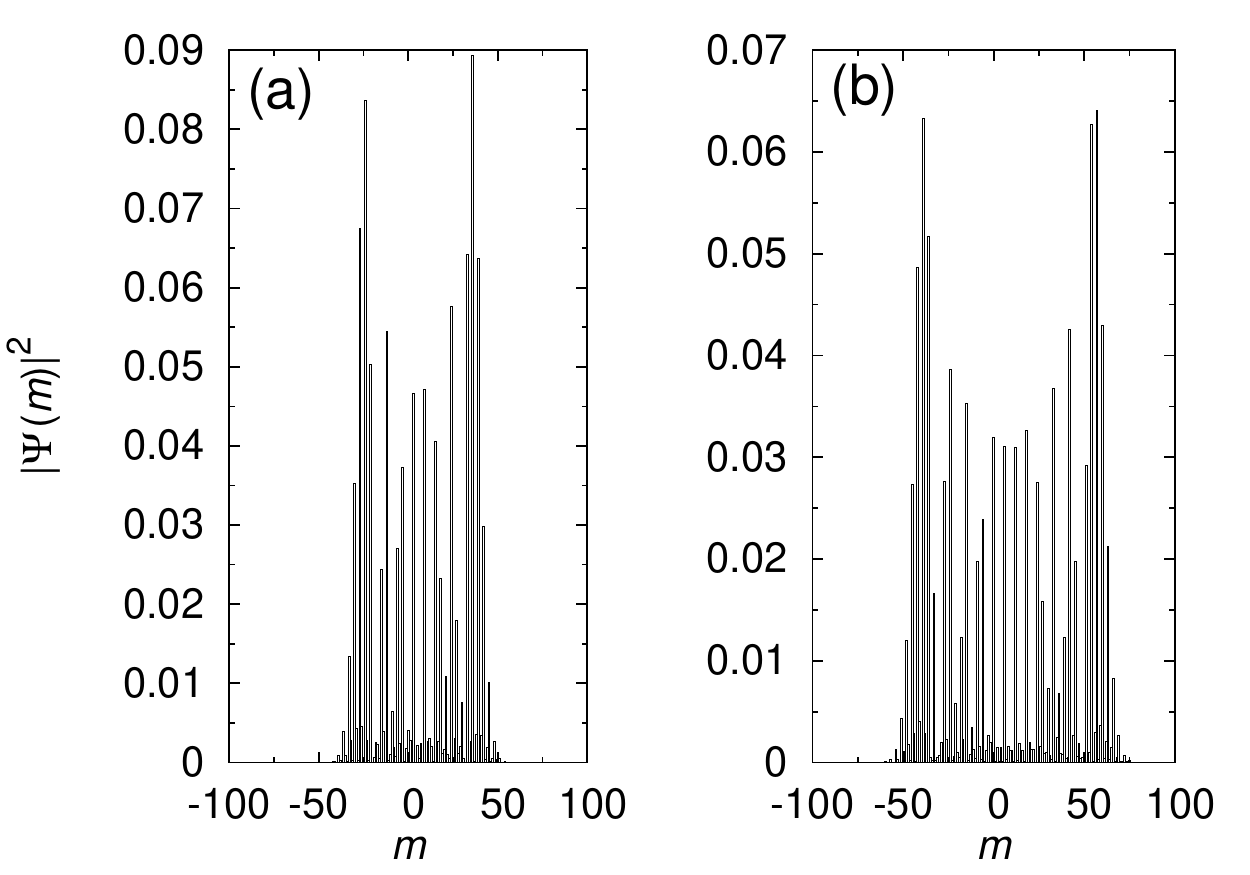}

\caption{\small{Final momentum distribution $ \left| \ip{m}{\Psi} \right|^{2}$ after (a) two 100-period adiabatic cycles and (b) after three 100-period adiabatic cycles. The initial state was the Wannier state $ \ket{\Psi_{3} (\alpha = 0)}$ for $ \eff{K} = 2 \heff $ (same as that used in Fig. 3(a) of the main text).}}
\end{figure}

% Numbers shown in (a) are the phases of each momentum component relative to the component $m=0$.

%\vspace{3cm}

%\begin{thebibliography}{99}
%\end{thebibliography}

\end{document}